\documentclass[12pt]{article}
\usepackage{a4wide}
\usepackage{latexsym}
\usepackage{amsmath}
\usepackage{amsfonts}
\usepackage{amscd}
\usepackage{shuffle}
\usepackage{cite}
\usepackage{graphicx}
\usepackage{float}

\usepackage{pslatex}
\usepackage[latin1]{inputenc}
\usepackage[OT2,T1]{fontenc}

\usepackage{ifthen}

\newcommand{\bq}{\begin{eqnarray}}
\newcommand{\eq}{\end{eqnarray}}
\newcommand{\eps}{\varepsilon}

\newboolean{convention_12n}
\setboolean{convention_12n}{false}

\begin{document}

\thispagestyle{empty}

\begin{flushright}
  MITP/15-064
\end{flushright}

\vspace{1.5cm}

\begin{center}
  {\Large\bf The CHY representation of tree-level primitive QCD amplitudes\\
  }
  \vspace{1cm}
  {\large Leonardo de la Cruz, Alexander Kniss and Stefan Weinzierl\\
\vspace{2mm}
      {\small \em PRISMA Cluster of Excellence, Institut f{\"u}r Physik, }\\
      {\small \em Johannes Gutenberg-Universit{\"a}t Mainz,}\\
      {\small \em D - 55099 Mainz, Germany}\\
  } 
\end{center}

\vspace{2cm}

\begin{abstract}\noindent
  {
In this paper we construct a CHY representation for all tree-level primitive QCD amplitudes.
The quarks may be massless or massive.
We define a generalised cyclic factor $\hat{C}(w,z)$ and a generalised permutation invariant function $\hat{E}(z,p,\eps)$.
The amplitude is then given as a contour integral encircling the solutions of the scattering equations
with the product $\hat{C} \hat{E}$ as integrand.
Equivalently, it is given as a sum over the inequivalent solutions of the scattering equations, where the summand 
consists of a Jacobian times the product $\hat{C} \hat{E}$.
This representation separates information:
The generalised cyclic factor does not depend on the helicities of the external particles, 
the generalised permutation invariant function does not depend on the ordering of the external particles.
   }
\end{abstract}

\vspace*{\fill}

\newpage

\section{Introduction}
\label{sect:intro}

The Cachazo-He-Yuang representation (CHY representation) of tree-level amplitudes is based on the solutions of the scattering equations. 
The scattering equations are a set of algebraic equations, which associate to the $n$ momentum vectors of a scattering event
$(n-3)!$ inequivalent $n$-tuples of complex numbers $z=(z_1,...,z_n)$.
These scattering equations have been studied in a series of papers by 
Cachazo, He and Yuang \cite{Cachazo:2013iaa,Cachazo:2013gna,Cachazo:2013hca,Cachazo:2013iea,Cachazo:2014nsa,Cachazo:2014xea,Cachazo:2015ksa}.
It is remarkable, that tree amplitudes for gluons (spin $1$) or gravitons (spin $2$)
can be expressed elegantly 
either as a contour integral localised at the zeros of the scattering equations
or equivalently as a sum over the $(n-3)!$ inequivalent solutions of the scattering equations.
The essential ingredients for the gluon amplitudes are the Parke-Taylor factor $C(w,z)$, defining the cyclic order and
a permutation invariant function $E(z,p,\eps)$, containing the information on the helicities of the external particles.
In the gluon case, the permutation invariant function $E(z,p,\eps)$ can be written as a (reduced) Pfaffian.
The CHY representation has triggered significant interest in the 
community \cite{Dolan:2013isa,Dolan:2014ega,He:2014wua,Naculich:2014naa,Naculich:2015zha,Naculich:2015coa,Weinzierl:2014ava,Kalousios:2013eca,Weinzierl:2014vwa,Lam:2014tga,Monteiro:2013rya,Cachazo:2015nwa,Baadsgaard:2015voa,Baadsgaard:2015hia,Baadsgaard:2015ifa}.
In addition, there are interesting connections with string theory \cite{Bjerrum-Bohr:2014qwa,Mason:2013sva,Berkovits:2013xba,Gomez:2013wza,Adamo:2013tsa,Geyer:2014fka,Casali:2014hfa,Geyer:2015bja}
and gravity \cite{Schwab:2014xua,Afkhami-Jeddi:2014fia,Zlotnikov:2014sva,Kalousios:2014uva,White:2014qia,Monteiro:2014cda}.

The CHY representation of the tree-level gluon amplitudes separates information:
The Parke-Taylor factor does not depend on the helicities of the external particles, 
the permutation invariant function does not depend on the ordering of the external particles.
We may ask, if this separation of information exists for other cyclic ordered amplitudes.
It is known that this is the case for 
tree amplitudes in ${\mathcal N}=4$ super-Yang-Mills theories (SYM) and for tree amplitudes in QCD with one massless quark-anti-quark pair
and an arbitrary number of gluons \cite{Weinzierl:2014ava}.
These amplitudes satisfy as the pure gluon amplitudes cyclic invariance, the Kleiss-Kuijf relations \cite{Kleiss:1988ne}
and the Bern-Carrasco-Johansson relations (BCJ relations) \cite{Bern:2008qj}.
These relations among amplitudes with different cyclic order are encoded within the CHY representation in the Parke-Taylor factor.
As a consequence, amplitudes in ${\mathcal N}=4$ SYM and QCD amplitudes with one quark-anti-quark pair have a CHY representation
with the same Parke-Taylor factor $C(w,z)$ and a modified permutation invariant function $\hat{E}(z,p,\eps)$.
The situation is more complicated for tree-level primitive QCD amplitudes with more than one quark-anti-quark pair.
These amplitudes do not satisfy the full set of BCJ relations and the cyclic order can therefore not be represented by the standard
Parke-Taylor factor.
These amplitudes will require in addition to the generalisation $\hat{E}(z,p,\eps)$ of the permutation invariant function a generalisation
$\hat{C}(w,z)$ of the standard Parke-Taylor factor.
For the simplest case of the QCD tree-level four-point amplitude $A_4(\bar{q},q,\bar{q}',q')$ with two quark-anti-quark pairs 
this has been discussed in \cite{Weinzierl:2014ava}.
However, what is still missing is a treatment of an arbitrary tree-level primitive QCD amplitude.
In order to construct a CHY representation for these amplitudes, we need to know the relations among the amplitudes with
different external orderings.
Quite recently it was discovered that tree-level primitive QCD amplitude satisfy apart from some well-known 
``no-crossed-fermion-lines''-relations a well-described restricted set of fundamental BCJ relations.
This was first conjectured in \cite{Johansson:2015oia} and subsequently proven in \cite{delaCruz:2015dpa}.
Equipped with this information one may first construct a minimal basis for the amplitudes, and as the number of the elements
of the minimal basis never exceeds $(n-3)!$ construct a CHY representation. This is the content of this paper.

In this paper we show that all tree-level primitive QCD amplitudes have a representation of the form
\bq
\label{intro_integral_representation}
 A_n\left(w,p,\eps\right)
 & = &
 \frac{i}{\left(2\pi i\right)^{n-3}}
 \int
 \frac{d^nz}{d\omega}
 \;
 \prod{}' \delta\left( f_a\left(z,p\right) \right)
 \;
 \hat{C}\left(w, z\right) 
 \; 
 \hat{E}\left(z,p,\eps\right),
\eq
or equivalently
\bq
\label{intro_sum_representation}
 A_n\left(w,p,\eps\right)
 & = &
 i
 \sum\limits_{\mathrm{solutions} \; j} J\left(z^{(j)},p\right) \; \hat{C}\left(w, z^{(j)}\right) \; \hat{E}\left(z^{(j)},p,\eps\right).
\eq
The precise definition of all quantities will be given later on in the main text:
The arguments of the amplitudes on the left-hand side will be defined in section~\ref{sect_basic_definitions},
the integral measure of eq.~(\ref{intro_integral_representation}) will be defined in section~\ref{sect:CHY_representation_QCD} and
the Jacobian factor $J(z,p)$ appearing in eq.~(\ref{intro_sum_representation}) will be defined in section~\ref{sect:jacobian}.
Central to this paper are the generalised cyclic factor $\hat{C}(w,z)$ and the generalised permutation invariant
function $\hat{E}(z,p,\eps)$, which will treated in section~\ref{sect:def_C_hat} and section~\ref{sect:def_E_hat}, respectively.

Note that tree-level amplitudes in any theory defined by a Lagrangian can be computed easily by a variety of methods (Feynman diagrams,
Berends-Giele recursion relations \cite{Berends:1987me}, 
BCFW-recursion relations \cite{Britto:2005fq}) and can be considered as known quantities.
The purpose of this paper is to show that these quantities have a representation 
in the form of eq.~(\ref{intro_integral_representation}) or eq.~(\ref{intro_sum_representation})
and to provide a definition of the generalised cyclic factor $\hat{C}(w,z)$
and the generalised permutation invariant function $\hat{E}(z,p,\eps)$.
The virtue of a representation in the form of eq.~(\ref{intro_integral_representation}) or eq.~(\ref{intro_sum_representation})
lies in the fact that it separates the information on the external ordering (contained in the generalised cyclic factor $\hat{C}(w,z)$)
from the information on the helicities of the external particles (contained in the generalised permutation invariant function 
$\hat{E}(z,p,\eps)$).

Our construction relies on one conjecture.
The conjecture is stated in eq.~(\ref{rank_condition_F}).
In simple terms, the conjecture says that the external orderings of a minimal amplitude basis for $n_q>0$ 
remain linearly independent, when viewed as the external orderings 
of the pure gluonic ($n_q=0$) amplitudes.
We have verified this conjecture for all amplitudes up to 10 points.

This paper is organised as follows: 
In section~\ref{sect:primitive} we review basic facts about tree-level primitive QCD amplitudes.
It will be convenient to introduce words and shuffle algebras.
We summarise the relations among the primitive amplitudes and define a basis of primitive amplitudes.
In section~\ref{sect:scattering_equations} we introduce the scattering equations.
Since we are interested in primitive QCD amplitudes with massless or massive quarks, we present the extension
of the scattering equations to the massive case for QCD amplitudes.
In this section we also define the Jacobian $J(z,p)$.
Section~\ref{sect:CHY_representation_QCD} contains the main result of this paper.
We define the generalised cyclic factor $\hat{C}(w,z)$ and the generalised permutation invariant function $\hat{E}(z,p,\eps)$.
We then prove that with these definitions all tree-level primitive QCD amplitudes agree with the CHY representation.
In order to illustrate our approach, we work out in section~\ref{sect:example} a non-trivial example.
Finally, section~\ref{sect:conclusions} contains our conclusions.
In an appendix we collected a few technical details:
The proof of an equation allowing the orientation of fermion lines (appendix~\ref{sect:fermion_orientation}),
the explicit expressions of the coefficients appearing in the general BCJ relation (appendix~\ref{sect:def_F_w_wp})
and a proof that a weaker statement is sufficient to prove the above-mentioned conjecture (appendix~\ref{sect:upper_triangle_block}).

\section{Tree-level primitive QCD amplitudes}
\label{sect:primitive}

In this section we introduce our notation. We define words and shuffle algebras
and review the various relations among primitive amplitudes.
Dyck words are a convenient tool to label amplitudes with several quark-anti-quark pairs.
At the end of this section we present a minimal amplitude basis.

\subsection{Basic definitions}
\label{sect_basic_definitions}

Let us consider a tree-level primitive QCD amplitude with $n$ external particles, 
out of which $n_q$ particles are quarks, $n_q$ particles are anti-quarks 
and $n_g$ particles are gluons \cite{Bern:1994fz,Reuschle:2013qna}. 
We have the obvious relation
\bq
 n & = & n_g + 2 n_q.
\eq
Without loss of generality we may assume that all quarks have different flavours.
The quarks may be massless or massive.
We label the quarks by $q_1, q_2, ..., q_{n_q}$,
the corresponding anti-quarks by $\bar{q}_1, \bar{q}_2, ..., \bar{q}_{n_q}$,
and the gluons by $g_1, g_2, ..., g_{n_g}$.
We call the set
\bq
 {\mathbb A}
 & = & \left\{q_1, q_2, ..., q_{n_q}, \bar{q}_1, \bar{q}_2, ..., \bar{q}_{n_q}, g_1, g_2, ..., g_{n_g} \right\}
\eq
an alphabet and the elements of this set letters.
Ordered sequences of letters are called words:
\bq
 w & = & l_1 l_2 ... l_n.
\eq
We are in particular interested in words with $n$ letters, such that every letter from the alphabet
occurs exactly once.
We denote the set of these words by
\bq
 W_0
 & = &
 \left\{ \; l_1 l_2 ... l_n \; | \; l_i \in {\mathbb A}, \; l_i \neq l_j \;\mbox{for} \; i \neq j \; \right\}.
\eq
The set $W_0$ has $n!$ elements and each element of $W_0$ can be considered as a permutation of the $n$ letters of the alphabet ${\mathbb A}$.
For later purpose we define the reversed word $w^T$ by
\bq
\label{reversed_word}
 w^T
 & = &
 l_n ... l_2 l_1.
\eq
The word of length zero is denoted by $e$.
The words from an alphabet form an algebra.
The shuffle product $\shuffle$ of two words $w_1=l_1 l_2 ... l_k$ and $w_2 = l_{k+1} ... l_r$ 
is defined by
\bq
\label{def_shuffle}
 l_1 l_2 ... l_k \; \shuffle \; l_{k+1} ... l_r 
 & = &
 \sum\limits_{\mbox{\tiny shuffles} \; \sigma} l_{\sigma(1)} l_{\sigma(2)} ... l_{\sigma(r)},
\eq
where the sum runs over all permutations $\sigma$, which preserve the relative order of $l_1,l_2,...,l_k$ and of $l_{k+1},...,l_r$.
The shuffle product is commutative and associative:
\bq
 w_1 \shuffle w_2 & = & w_2 \shuffle w_1,
 \nonumber \\
 \left( w_1 \shuffle w_2 \right) \shuffle w_3 & = & w_1 \shuffle \left( w_2 \shuffle w_3 \right).
\eq
The name ``ordered permutations'' is also used for the shuffle product.
The empty word $e$ is the unit in this algebra:
\bq
 e \shuffle w = w \shuffle e = w.
\eq
We can use the words $w \in W_0$ to encode the order of the external particles of tree-level primitive QCD amplitudes
and we will write
\bq
\label{short_notation}
 A_n\left(w\right)
 & \mbox{or} &
 A_n\left( l_1 l_2 ... l_n \right)
\eq
for such an amplitude. The external momenta for this amplitude are denoted by $p_1$, $p_2$, ..., $p_n$.
The $n$-tuple of external momenta will be denoted by $p=(p_1,...,p_n)$.
In a similar way we will denote the $n$-tuple of external polarisations by $\eps$.
The external polarisations are given by polarisation vectors $\eps_j$ for external gluons,
spinors $\bar{u}_j$ for out-going fermions and spinors $v_j$ for out-going anti-fermions.
For simplicity we will assume all particles to be out-going.
We will write
\bq
 A_n\left(w,p,\eps\right)
\eq
if we would like to emphasize that the primitive amplitude depends apart from the external ordering $w$ also
on the external momenta $p$ and the polarisations $\eps$.
In situations, where the main focus is on the dependence on $w$, we will simply write $A_n(w)$ as in eq.~(\ref{short_notation}).
It will be convenient to introduce the following notation:
If $\lambda_1, \lambda_2$ are numbers and $w_1, w_2 \in W_0$ words, we write
\bq
\label{linear_operator_notation}
 A_n\left( \lambda_1 w_1 + \lambda_2 w_2 \right)
\eq
for
\bq
 \lambda_1 A_n\left(w_1\right)
 +
 \lambda_2 A_n\left(w_2\right).
\eq
In other words, we take $A_n$ as a linear operator on the vector space of words with basis $W_0$.
We will use this notation as a convenient way to express relations among primitive amplitudes.

\subsection{Relations among primitive amplitudes}
\label{sect:relations}

The primitive amplitudes are cyclic invariant:
\bq
\label{cyclic_invariance}
 A_n\left( l_1 l_2 ... l_n \right)
 & = &
 A_n\left( l_2 ... l_n l_1 \right).
\eq
Eq.~(\ref{cyclic_invariance}) is a first (and trivial) example of relations
among primitive amplitudes with different external ordering.
There are more relations among primitive amplitudes.
A further example are the Kleiss-Kuijf relations \cite{Kleiss:1988ne}.
Let
\bq
 w_1 = l_{\alpha_1} l_{\alpha_2} ... l_{\alpha_j},
 & & 
 w_2 = l_{\beta_1} l_{\beta_2} ... l_{\beta_{n-2-j}}
\eq
be two sub-words, such that
\bq
 \{ l_1 \} \cup 
 \{ l_{\alpha_1}, ..., l_{\alpha_j}  \} \cup \{ l_{\beta_1}, ..., l_{\beta_{n-2-j}} \}
 \cup \{ l_n \}
 & = &
 \{ l_1,...,l_{n} \}.
\eq
Then
\bq
\label{Kleiss_Kuijf}
 A_n\left( l_1 l_{\alpha_1} ... l_{\alpha_j} l_{n} l_{\beta_1} ... l_{\beta_{n-2-j}} \right)
 & = & 
 \left( -1 \right)^{n-2-j}
 A_n\left( \; l_1 \left( w_1 \shuffle w_2^T \right) l_{n} \; \right).
\eq
We recall that $w^T$ denotes the reversed word, defined in eq.~(\ref{reversed_word}),
the symbol $\shuffle$ denotes the shuffle product, defined in eq.~(\ref{def_shuffle})
and we used the notation of eq.~(\ref{linear_operator_notation}).
The Kleiss-Kuijf relations in eq.~(\ref{Kleiss_Kuijf}) allow us to fix two legs at specified positions.

A special case of the Kleiss-Kuijf relation is the case, where $w_1$ is the empty word.
In this case the Kleiss-Kuijf relation reduces to the reflection identity for primitive amplitudes
\bq
\label{reflection_identity}
 A_n\left( w \right)
 & = & 
 \left( -1 \right)^{n}
 A_n\left( w^T \right).
\eq
A second special case is given for the situation, where the set $\beta$ contains only one element. 
In this case the Kleiss-Kuijf relation reduces to the $U(1)$-decoupling identity
\bq
\label{U1_decoupling}
 \sum\limits_{\sigma \in {\mathbb Z}_{n-1}}
 A_n\left(l_{\sigma_1} l_{\sigma_2} ... l_{\sigma_{n-1}} l_n\right) 
 & = & 0,
\eq
where the sum is over the cyclic permutations of the first $(n-1)$ arguments.

For amplitudes with more than one quark line ($n_q > 1$) there are some trivial relations related to the fact that 
primitive amplitudes cannot have crossed fermion lines.
Tree-level primitive amplitudes have a fixed cyclic order and all Feynman diagrams contributing 
to such an amplitude can be drawn in a planar way on a disc.
If the amplitude has crossed fermion lines the diagrams can only be drawn in a planar way with 
flavour-changing currents.
However, in QCD there are no flavour-changing currents and these amplitudes are zero.
Thus we have the relations:
\bq
\label{no_crossed_fermions}
 A_n\left( ... q_i ... q_j ... \bar{q}_i ... \bar{q}_j ... \right)
 \;\; = \;\;
 A_n\left( ... q_i ... \bar{q}_j ... \bar{q}_i ... q_j ... \right)
 \;\; = \;\;
 0.
\eq
For amplitudes with at least one gluon there are further relations.
Let us assume that particle $2$ is a gluon:
\bq
 l_2 & = & g_\alpha,
 \;\;\;\;\;\;
 \alpha \in \{1,...,n_g\}.
\eq
The fundamental Bern-Carrasco-Johansson relations (BCJ relations) read
\bq
\label{fundamental_BCJ_relation}
 \sum\limits_{i=2}^{n-1} 
  \left( \sum\limits_{j=i+1}^n 2 p_2 p_j \right)
  A_n\left( l_1 l_3 ... l_i l_2 l_{i+1} ... l_{n-1} l_n \right)
 & = & 0.
\eq
These relations have first been conjectured for pure gluon amplitudes \cite{Bern:2008qj}
and proven in this case in \cite{BjerrumBohr:2009rd,Stieberger:2009hq,Feng:2010my}.
The conjecture was later extended to all tree-level primitive QCD amplitudes \cite{Johansson:2015oia}
and proven in \cite{delaCruz:2015dpa}.

Let us summarise: The relations among tree-level primitive QCD amplitudes are
\begin{enumerate}
\item Cyclic invariance, stated in eq.~(\ref{cyclic_invariance}),
\item the Kleiss-Kuijf relations, given in eq.~(\ref{Kleiss_Kuijf}),
\item the ``no-crossed-fermion-lines''-relation in eq.~(\ref{no_crossed_fermions}),
\item the fundamental BCJ relations stated in eq.~(\ref{fundamental_BCJ_relation}).
\end{enumerate}

\subsection{Dyck words}
\label{sect:Dyck}

Primitive amplitudes with crossed fermion lines vanish.
The ones with no crossed fermion lines may be described by generalised Dyck words \cite{Melia:2013bta,Melia:2013epa}.
In order to define these generalised Dyck let us consider an alphabet consisting
of $n_q$ distinct opening brackets ``$(_i$'' and $n_q$ corresponding closing brackets ``$)_i$''.
Closing brackets of type $i$ only match with opening brackets of type $i$.
A generalised Dyck word is any word from this alphabet with properly matched brackets.
Originally, Dyck did not consider brackets of different types.
We will use the term ``Dyck word'' if there is only one type of brackets and the term
``generalised Dyck word'' in the case of brackets with more than one type.
We are mainly interested in the generalised Dyck words of length $2n_q$, where every opening and
every closing bracket occurs exactly once.
There are 
\bq
 N_{\mathrm{Dyck}}
 & = &
 \frac{\left(2 n_q\right)!}{\left(n_q+1\right)!}
\eq
words of this type.
The opening and the closing brackets of type $i$ may be associated to the fermion line $i$.
There are two possible orientations for each fermion line, either
\bq
\label{def_standard_orientation}
 q_i \rightarrow (_i,
 & &
 \bar{q}_i \rightarrow )_i,
\eq
or
\bq
 \bar{q}_i \rightarrow (_i,
 & &
 q_i \rightarrow )_i.
\eq
We define a standard orientation of the fermion lines 
by requiring, that every quark corresponds to an opening bracket
and every anti-quark corresponds to a closing bracket,
i.e. the standard orientation is given for each fermion line by eq.~(\ref{def_standard_orientation}).
This definition is not cyclic invariant, 
however we may always use the Kleiss-Kuijf relations to fix particle $1$ to be $q_1$ and particle $n$
to be $\bar{q}_1$.
Let us define a projection $P$ by
\bq
 P\left(q_i\right) \;\; = \;\; (_i,
 \;\;\;\;\;\;
 P\left(g_i\right) \;\; = \;\; e,
 \;\;\;\;\;\;
 P\left(\bar{q}_i\right) \;\; = \;\; )_i.
\eq
We then set
\bq
 \mathrm{Dyck}_{n_q}
 & = &
 \left\{ \; w \in W_0 \; | \; P(w) \; \mbox{is a generalised Dyck word} \; \right\}.
\eq
This set contains all words without crossed fermion lines and where all fermion lines have the
standard orientation.

It is always possible to reduce an amplitude with an arbitrary orientation of the fermion lines 
to the standard orientation of the fermion lines, by just using cyclic invariance, the Kleiss-Kuijf relations
and the ``no-crossed-fermion-lines''-relations \cite{Melia:2013bta,Melia:2013epa}.
In order to see this, let us assign for amplitudes with no crossed fermion lines 
a level to each fermion line.
We draw the external order of the particles on the boundary of a disc and we draw on the disc for each
quark-anti-quark-pair a fermion line connecting the anti-quark with the corresponding quark.
With the help of the Kleiss-Kuijf relations we may always put the quark $q_1$ at position $1$ and
the corresponding anti-quark $\bar{q}_1$ at position $n$.
We assign level $0$ to this fermion line.
We assign level $1$ to all fermion lines, which are not separated by another fermion line from the
fermion line of level $0$.
We then iterate this procedure and we assign level $k$ to all fermions line, which are not 
separated by another fermion line from some fermion line of level $(k-1)$, and which have not been
assigned any level before.

There is an iterative procedure, which allows us to express an amplitude 
with an arbitrary orientation of the fermion lines as a linear combination of amplitudes with the
standard orientation.
This procedure brings first all fermion lines of level $1$ into the standard orientation, then
all fermion lines of level $2$, etc..
The fermion line of level $0$ is trivially brought into the standard orientation with the help
of the Kleiss-Kuijf relations.
At level $k$ consider the amplitude
\bq
 A_n\left( x_{k-1} q_i x_k \bar{q}_j w_{k+1} q_j y_k \bar{q}_i y_{k-1} \right),
\eq
where $x_{k-1}$, $x_k$, $w_{k+1}$, $y_k$ and $y_{k-1}$ are sub-words.
We assume that the fermion line $q_i$-$\bar{q}_i$ is of level $(k-1)$.
This fermion line has already the standard orientation and we assume that all fermion lines
contained in the sub-words $x_{k-1}$ and $y_{k-1}$ have already been oriented.
The fermion line $q_j$-$\bar{q}_j$ is of level $k$ and has the wrong orientation.
The sub-words $x_k$ and $y_k$ may contain further fermion lines of level $k$ and higher level.
The sub-word $w_{k+1}$ may contain fermion lines of level $(k+1)$ and higher.
We are going to orient the fermion line $q_j$-$\bar{q}_j$, respecting the orientations of all fermion lines with
level $\le k$.
Let us write
\bq
 x_k \;\; = \;\; l_{i_1} l_{i_2} ... l_{i_r},
 & &
 y_k \;\; = \;\; l_{j_1} l_{j_2} ... l_{j_s}.
\eq
Then 
\bq
\label{fermion_orientation}
\lefteqn{
 A_n\left( x_{k-1} q_i x_k \bar{q}_j w_{k+1} q_j y_k \bar{q}_i y_{k-1} \right)
 = } & &
 \nonumber \\
 & &
 \left(-1\right)^{|w_{k+1}|+1}
 \sum\limits_{a=0}^r
 \sum\limits_{b=0}^s
 A_n\left( x_{k-1} q_i l_{i_1} ... l_{i_a} q_j w_{k+1}' \bar{q}_j l_{j_{b+1}} ... l_{j_s} \bar{q}_i y_{k-1} \right),
\eq
where $|w_{k+1}|$ denotes the length of the sub-word $w_{k+1}$ and with
\bq
 w_{k+1}'
 & = &
 \left( l_{i_{a+1}} ... l_{i_r} \right) \shuffle w_{k+1}^T \shuffle \left( l_{j_1} ... l_{j_b} \right).
\eq
All fermion lines of $w_{k+1}'$ are of level $(k+1)$ or higher.
We call eq.~(\ref{fermion_orientation}) the ``fermion orientation'' relations.
Note that some amplitudes in eq.~(\ref{fermion_orientation}) may be zero due to crossed fermion lines.
This is either the case if a quark-anti-quark pair from $x_k$ is split between $l_{i_1} ... l_{i_a}$ and
$w_{k+1}'$ or if a quark-anti-quark pair from $y_k$ is split between $w_{k+1}'$ and $l_{j_{b+1}} ... l_{j_s}$.
We give a proof of eq.~(\ref{fermion_orientation}) in appendix~\ref{sect:fermion_orientation}.

\subsection{The amplitude basis}
\label{sect:amplitude_basis}

The relations among tree-level primitive QCD amplitudes allows us to express all amplitudes for a given set
of external particles in terms of a set of basis amplitudes.
The size of this basis is
\bq
 N_{\mathrm{basis}}
 & = &
 \left\{
 \begin{array}{ll}
   \left(n-3\right)!, & n_q \in \{0,1\}, \\
   \left(n-3\right)! \frac{2\left(n_q-1\right)}{n_q!}, & n_q \ge 2. \\
 \end{array}
 \right.
\eq
For later purpose we set
\bq
 N_{\mathrm{solutions}}
 & = &
 \left(n-3\right)!,
\eq
(the subscript is a reminder that $(n-3)!$ is the number of inequivalent solutions of the scattering equations)
and
\bq
 N_{\mathrm{permutations}}
 & = &
 n!.
\eq
Note that 
\bq
 \frac{2\left(n_q-1\right)}{n_q!}
 \;\; = \;\;
 \frac{2}{n_q} \frac{1}{\left(n_q-2\right)!} 
 \;\; \le \;\;
 1,
 \;\;\;\;\;\;
 \mbox{for} \;\; n_q \ge 2,
\eq
and therefore we always have
\bq
\label{inequality}
 N_{\mathrm{basis}}
 & \le &
 N_{\mathrm{solutions}}.
\eq
In order to find a CHY representation for tree-level primitive QCD amplitudes it is essential
that the number of basis amplitudes 
does not exceed the number of inequivalent solutions of the scattering equations.
Eq.~(\ref{inequality}) shows that this condition is always satisfied.

Let us now describe the amplitude basis for the various cases.
For $n_q=0$ the set of words corresponding to a possible basis is given by \cite{Johansson:2015oia}
\ifthenelse{\boolean{convention_12n}}
{
\bq
\label{basis_n_q_0}
 B 
 & = &
 \left\{ 
  \; l_1 l_2 ... l_n \in W_0 \; | \; l_{1}=g_{1}, \; l_{2}=g_{2}, \; l_n=g_n \;
 \right\}.
\eq
}
{
\bq
\label{basis_n_q_0}
 B 
 & = &
 \left\{ 
  \; l_1 l_2 ... l_n \in W_0 \; | \; l_{1}=g_{1}, \; l_{n-1}=g_{n-1}, \; l_n=g_n \;
 \right\}.
\eq
}
For $n_q=1$ we may choose
\ifthenelse{\boolean{convention_12n}}
{
\bq
\label{basis_n_q_1}
 B 
 & = &
 \left\{ 
  \; l_1 l_2 ... l_n \in W_0 \; | \; l_{1}=q_{1}, \; l_{2}=g_{1}, \; l_n=\bar{q}_1 \;
 \right\}.
\eq
}
{
\bq
\label{basis_n_q_1}
 B 
 & = &
 \left\{ 
  \; l_1 l_2 ... l_n \in W_0 \; | \; l_{1}=q_{1}, \; l_{n-1}=g_{n-2}, \; l_n=\bar{q}_1 \;
 \right\}.
\eq
}
For $n_q\ge 2$ we may choose
\ifthenelse{\boolean{convention_12n}}
{
\bq
\label{basis_n_q_2}
 B 
 & = &
 \left\{ 
  \; l_1 l_2 ... l_n \in \mathrm{Dyck}_{n_q} \; | \; l_{1}=q_1, \; l_{2} \in \{q_2,...,q_{n_q}\}, \; l_n=\bar{q}_1 \;
 \right\}.
\eq
}
{
\bq
\label{basis_n_q_2}
 B 
 & = &
 \left\{ 
  \; l_1 l_2 ... l_n \in \mathrm{Dyck}_{n_q} \; | \; l_{1}=q_1, \; l_{n-1} \in \{\bar{q}_2,...,\bar{q}_{n_q}\}, \; l_n=\bar{q}_1 \;
 \right\}.
\eq
}
Let us briefly review how to express an arbitrary amplitude $A_n(w)$ with $w \in W_0$ as 
a linear combination of amplitudes $A_n(w_j)$ with $w_j \in B$, using the relations
summarised in section~\ref{sect:relations}.

We first use cyclic invariance as in eq.~(\ref{cyclic_invariance})
to fix particle $1$ to be $g_1$ (in the pure gluonic case $n_q=0$) or
to be $q_1$ (in the case $n_q\ge 1$).
Let us define a subset $W_1$ of $W_0$ by
\bq
 W_1 & = &
 \left\{
 \begin{array}{ll}
  \left\{ \; l_1 l_2 ... l_n \in W_0 \; | \; l_1=g_1 \; \right\}, & n_q=0, \\
  \left\{ \; l_1 l_2 ... l_n \in W_0 \; | \; l_1=q_1 \; \right\}, & n_q \ge 1. \\
 \end{array}
 \right.
\eq
The set $W_1$ contains all words, where the first letter has been fixed.
We then use the Kleiss-Kuijf relations in eq.~(\ref{Kleiss_Kuijf}) 
to fix particle $n$ to be $g_n$ (in the pure gluonic case $n_q=0$) or
to be $\bar{q}_1$ (in the case $n_q\ge 1$).
We define a subset $W_2$ of $W_1$ by
\bq
 W_2 & = &
 \left\{
 \begin{array}{ll}
  \left\{ \; l_1 l_2 ... l_n \in W_1 \; | \; l_n=g_n \; \right\}, & n_q=0, \\
  \left\{ \; l_1 l_2 ... l_n \in W_1 \; | \; l_n=\bar{q}_1 \; \right\}, & n_q \ge 1. \\
 \end{array}
 \right.
\eq
The set $W_2$ contains all words, where the first and the last letter have been fixed.
If $n_q\ge2$ we then set to zero any amplitude with crossed fermion lines, in accordance with eq.~(\ref{no_crossed_fermions}).
We then use eq.~(\ref{fermion_orientation})
to express amplitudes with no crossed fermion lines 
in terms of amplitudes with no crossed fermion lines 
and the standard orientation of the fermion lines.
The standard orientation of the fermion lines has been defined in eq.~(\ref{def_standard_orientation}).
We define a subset $W_3$ of $W_2$ by
\bq
 W_3 & = &
 \left\{
 \begin{array}{ll}
  W_2, & n_q \le 1, \\
  \left\{ \; w \in W_2 \; | \; w \in \mathrm{Dyck}_{n_q} \; \right\}, & n_q \ge 2. \\
 \end{array}
 \right.
\eq
The set $W_3$ contains all words, where the first and the last letter have been fixed.
In addition $W_3$ excludes all words, which either correspond to crossed fermion lines or correspond to a 
non-standard orientation of the fermion lines.
\ifthenelse{\boolean{convention_12n}}
{Finally, we use the fundamental BCJ relation of eq.~(\ref{fundamental_BCJ_relation})
to fix particle $2$ 
to be $g_{2}$ (in the pure gluonic case $n_q=0$),
to be $g_{1}$ (in the case $n_q=1$) or
to remove any gluon from position $2$ (in the case $n_q\ge 2$).
In the latter case we then have necessarily a quark at position $2$, as we already have chosen the standard orientation.}
{Finally, we use the fundamental BCJ relation of eq.~(\ref{fundamental_BCJ_relation})
to fix particle $(n-1)$ 
to be $g_{n-1}$ (in the pure gluonic case $n_q=0$),
to be $g_{n-2}$ (in the case $n_q=1$) or
to remove any gluon from position $(n-1)$ (in the case $n_q\ge 2$).
In the latter case we then have necessarily an anti-quark at position $(n-1)$, as we already have chosen the standard orientation.}
This brings us down to the basis
\ifthenelse{\boolean{convention_12n}}
{
\bq
 B & = &
 \left\{
 \begin{array}{ll}
  \left\{ \; l_1 l_2 ... l_n \in W_3 \; | \; l_{2}=g_{2} \; \right\}, & n_q=0, \\
  \left\{ \; l_1 l_2 ... l_n \in W_3 \; | \; l_{2}=g_{1} \; \right\}, & n_q=1, \\
  \left\{ \; l_1 l_2 ... l_n \in W_3 \; | \; l_{2} \in \{q_2,...,q_{n_q}\} \; \right\}, & n_q \ge 2. \\
 \end{array}
 \right.
\eq
}
{
\bq
 B & = &
 \left\{
 \begin{array}{ll}
  \left\{ \; l_1 l_2 ... l_n \in W_3 \; | \; l_{n-1}=g_{n-1} \; \right\}, & n_q=0, \\
  \left\{ \; l_1 l_2 ... l_n \in W_3 \; | \; l_{n-1}=g_{n-2} \; \right\}, & n_q=1, \\
  \left\{ \; l_1 l_2 ... l_n \in W_3 \; | \; l_{n-1} \in \{\bar{q}_2,...,\bar{q}_{n_q}\} \; \right\}, & n_q \ge 2. \\
 \end{array}
 \right.
\eq
}
The set $B$ contains all words corresponding to a possible basis, 
as already stated in eqs.~(\ref{basis_n_q_0})-~(\ref{basis_n_q_2}).
We have the inclusions
\bq
 W_0 \supseteq W_1 \supseteq W_2 \supseteq W_3 \supseteq B.
\eq
We will use this chain of inclusions for constructions and proofs in this paper.

We already mentioned that we may view $A_n$ as a linear operator on the vector space of words with basis $W_0$. Let us denote this vector space by $V$.
The dimension of $V$ is $N_{\mathrm{permutations}}=n!$.
Let us assume, that there is another linear operator $\tilde{A}_n$ on $V$.
We would like to investigate, under which conditions $A_n$ and $\tilde{A}_n$ are identical.
This is the case if and only if they agree on all basis vectors of $V$:
\bq
\label{identical_linear_operators}
 \tilde{A}_n\left(w\right)
 & = &
 A_n\left(w\right),
 \;\;\;\;\;\;
 \forall w \in W_0.
\eq
However, we further know that there are relations among the $A_n(w_j)$, and if $A_n$ and $\tilde{A}_n$
are identical operators, we must have the same relations among the $\tilde{A}_n(w_j)$.
Therefore it is sufficient to check that $\tilde{A}_n$ and $A_n$ agree on the smaller set $B$
and to check that the images $\tilde{A}_n(w_j)$ satisfy all the relations of section~\ref{sect:relations}.
Actually it is sufficient to check, that 
\begin{enumerate}
\item \label{cond1} $\tilde{A}_n(w)$ satisfies for all $w \in W_0$ cyclic invariance, stated in eq.~(\ref{cyclic_invariance}).
\item \label{cond2} $\tilde{A}_n(w)$ satisfies for all $w \in W_1$ the Kleiss-Kuijf relations of eq.~(\ref{Kleiss_Kuijf}).
\item \label{cond3} $\tilde{A}_n(w)$ satisfies for all $w \in W_2$ the ``no-crossed-fermion-lines''-relations of eq.~(\ref{no_crossed_fermions})
and the fermion orientation relations of eq.~(\ref{fermion_orientation}).
\item \label{cond4} $\tilde{A}_n(w)$ satisfies for all $w \in W_3$ the fundamental BCJ relations of eq.~(\ref{fundamental_BCJ_relation}).
\item \label{cond5} $\tilde{A}_n(w)$ agrees for all $w \in B$ with $A_n$:
\bq
 \tilde{A}_n\left(w\right)
 & = &
 A_n\left(w\right),
 \;\;\;\;\;\;
 \forall w \in B.
\eq
\end{enumerate}
In order to see that these conditions are sufficient let us start with $w \in B$. 
Condition \ref{cond5} guarantees that $\tilde{A}_n(w)$ agrees with $A_n(w)$ on $B$.
Let's then move to $w \in W_3 \backslash B$.
The fundamental BCJ relations of condition \ref{cond4} 
ensure, that $\tilde{A}_n(w)$ may be expressed as a linear combination of $\tilde{A}_n(w')$ with $w'\in B$.
The same relation holds for $A_n(w)$ with $\tilde{A}_n(w)$ substituted by $A_n(w)$ and $\tilde{A}_n(w')$ substituted by $A_n(w')$.
Since we already know that $\tilde{A}_n(w)$ agrees with $A_n(w)$ on $B$, 
we conclude that $\tilde{A}_n(w)$ agrees with $A_n(w)$ on $W_3$.
We may repeat this argumentation with condition \ref{cond3} and show that $\tilde{A}_n(w)$ agrees with $A_n(w)$ on $W_2$.
Condition \ref{cond2} allows us then to conclude that they agree on $W_1$ and finally condition \ref{cond1} ensures
that $\tilde{A}_n(w)$ agrees with $A_n(w)$ on $W_0$.

\section{The scattering equations}
\label{sect:scattering_equations}

In this section we introduce the scattering equations.
We first treat the massless case and proceed afterwards to the massive case.
We will also define the Jacobian $J(z,p)$, which we will need later on.

Let us denote by $\Phi_n$ the momentum configuration space of $n$ external particles:
\bq
 \Phi_n & = &
 \left\{ \left(p_1,p_2,...,p_n\right) \in \left({\mathbb C} M\right)^n | p_1+p_2+...+p_n=0, p_{g_j}^2 = 0, p_{q_j}^2 = p_{\bar{q}_j}^2 = m_{q_j}^2 \right\}.
\eq
In other words, a $n$-tuple $p=(p_1, p_2, ..., p_n)$ of momentum vectors belongs to $\Phi_n$ if this $n$-tuple satisfies momentum conservation
and the mass-shell conditions. For gluons we have $p_{g_j}^2 = 0$, while for quarks we have $p_{q_j}^2 = p_{\bar{q}_j}^2 = m_{q_j}^2$.
The quarks may be massive or massless, in the latter case we have $m_{q_j}=0$.
Note that a quark and an anti-quark of the same flavour have the same mass.

We further denote by $\hat{\mathbb C} = {\mathbb C} \cup \{\infty\}$.
The space $\hat{\mathbb C}$ is equivalent to the complex projective space ${\mathbb C}{\mathbb P}^1$.
For amplitudes with $n$ external particles we consider the space $\hat{\mathbb C}^n$.
Points in $\hat{\mathbb C}^n$ will be denoted by $z=(z_1,z_2,...,z_n)$.
We use the convention that $z$ without any index denotes an $n$-tuple.
We set for $1\le i \le n$
\bq
 f_i\left(z,p\right) & = & 
 \sum\limits_{j=1, j \neq i}^n \frac{ 2 p_i \cdot p_j + 2 \Delta_{ij}}{z_i - z_j}.
\eq
The quantity $\Delta_{ij}$ will be defined below.
Differences like in the denominator will occur often in this article and we use the abbreviation
\bq
 z_{ij} & = & z_i - z_j.
\eq

\subsection{The massless case}
\label{sect:scattering_eq_massless}

Let us start our discussion with the massless case, for which
\bq
 \Delta_{ij} & = & 0.
\eq
The scattering equations, originally proposed in the massless case, read \cite{Cachazo:2013hca}
\bq
\label{scattering_equations}
 f_i\left(z,p\right) & = & 0.
\eq
For a fixed $p \in \Phi_n$ a solution of the scattering equation is a point $z \in \hat{\mathbb C}^n$, such that the
scattering equations in eq.~(\ref{scattering_equations}) are satisfied.

The scattering equations are invariant under the projective special linear group
$\mathrm{PSL}(2,{\mathbb C})=\mathrm{SL}(2,{\mathbb C})/{\mathbb Z}_2$.
Here, ${\mathbb Z}_2$ is given by $\{ {\bf 1}, -{\bf 1} \}$, with ${\bf 1}$ denoting the $(2 \times 2)$-unit matrix.
Let
\bq
 g = \left(\begin{array}{cc} a & b \\ c & d \\ \end{array} \right) & \in & \mathrm{PSL}(2,{\mathbb C}).
\eq
Each $g \in \mathrm{PSL}(2,{\mathbb C})$ acts on a single $z_i \in \hat{\mathbb C}$ as follows:
\bq
 g \cdot z_i & = &
 \frac{a z_i + b}{c z_i + d}.
\eq
We further set
\bq
 g \cdot \left(z_1, z_2, ..., z_n \right)
 & = &
 \left(g \cdot z_1, g \cdot z_2, ..., g \cdot z_n \right).
\eq
If $(z_1,z_2, ..., z_n)$ is a solution of eq.~(\ref{scattering_equations}), then also
$(z_1',z_2', ..., z_n') = g \cdot (z_1,z_2, ..., z_n)$ is a solution.
We call two solutions which are related by a $\mathrm{PSL}(2,{\mathbb C})$-transformation equivalent solutions.
We are in particular interested in the set of all inequivalent solutions of the scattering equations.
As shown in \cite{Cachazo:2013iaa,Cachazo:2013gna}, there are $(n-3)!$ different solutions not related by a $\mathrm{PSL}(2,{\mathbb C})$-transformation.
We will denote a solution by 
\bq
 z^{(j)} & = & \left( z_1^{(j)}, ..., z_n^{(j)} \right)
\eq
and a sum over the $(n-3)!$ inequivalent solutions by
\bq
 \sum\limits_{\mathrm{solution}\;j}
\eq
The $n$ scattering equations in eq.~(\ref{scattering_equations}) are not independent, only $(n-3)$ of them are.
The M\"obius invariance implies the relations
\bq
\label{dependent_scattering_equations}
 \sum\limits_{j=1}^n f_j\left(z,p\right) = 0,
 \;\;\;\;\;\;
 \sum\limits_{j=1}^n z_j f_j\left(z,p\right) = 0,
 \;\;\;\;\;\;
 \sum\limits_{j=1}^n z_j^2 f_j\left(z,p\right) = 0.
\eq

\subsection{The massive case}
\label{sect:scattering_eq_massive}

The extension of the scattering equations to the massive case has been considered in \cite{Naculich:2014naa}.
In the massive case the scattering equations remain invariant under $\mathrm{PSL}(2,{\mathbb C})$ provided
\bq
\label{massive_cond_1}
 \sum\limits_{j=1,j \neq i}^n \Delta_{ij} & = & m_i^2.
\eq
The relations in eq.~(\ref{dependent_scattering_equations}) remain valid provided that the quantities $\Delta_{ij}$ satisfy in addition
\bq
\label{massive_cond_2}
 \Delta_{ij} & = & \Delta_{ji}.
\eq
Let us now consider primitive multi-quark amplitudes with $n_q$ quarks, $n_q$ anti-quarks and $n_g$ gluons.
We may assume that the flavours of all $n_q$ quarks are distinct.
In this case we have that to every external quark $q_a$ corresponds an external 
anti-quark $\bar{q}_a$ with the same mass $m_a$.
Eq.~(\ref{massive_cond_1}) and eq.~(\ref{massive_cond_2}) are satisfied if we set
\bq
\label{extension_massive_case}
 \Delta_{q_a \bar{q}_a} 
 \;\;\; = \;\;\;
 \Delta_{\bar{q}_a q_a} 
 \;\;\; = \;\;\;
 m_{q_a}^2
\eq
and $\Delta_{ij}=0$ in all other cases.

Eq.~(\ref{extension_massive_case}) is easily understood as follows: The massless scattering equations are valid in any space-time dimensions.
Starting from $D=4$ space-time dimensions, let us consider a theory in $D+n_q$ space-time dimensions
(one time dimension and $(D+n_q-1)$ spacial dimensions), 
where the quark of flavour $a$ carries in the $a$-th extra dimension a momentum component $m_{q_a}$
and the anti-quark of flavour $a$ carries in the $a$-th extra dimension the momentum component $(-m_{q_a})$.
We take the signature of the metric to be $(+,-,-,-,...)$.

\subsection{The Jacobian}
\label{sect:jacobian}

Let us define a $n \times n$-matrix $\Phi(z,p)$ with entries
\bq
 \Phi_{ab}\left(z,p\right)
 & = &
 \frac{\partial f_a\left(z,p\right)}{\partial z_b}
 \;\; = \;\;
 \left\{
 \begin{array}{cc}
 \frac{2 p_a \cdot p_b + 2 \Delta_{ab}}{z_{ab}^2} & a \neq b, \\
 - \sum\limits_{j=1, j \neq a}^n \frac{2 p_a \cdot p_j + 2 \Delta_{aj}}{z_{aj}^2} & a=b. \\
 \end{array}
 \right.
\eq
Let $\Phi^{ijk}_{rst}(z,p)$ denote the $(n-3)\times(n-3)$-matrix, where the rows $\{i,j,k\}$ and the columns $\{r,s,t\}$ have
been deleted.
We set
\bq
 \det{}' \; \Phi\left(z,p\right)
 & = &
 \left(-1\right)^{i+j+k+r+s+t}
 \frac{\left|\Phi^{ijk}_{rst}(z,p)\right|}{\left(z_{ij}z_{jk}z_{ki}\right)\left(z_{rs}z_{st}z_{tr}\right)}.
\eq
With the above sign included, the quantity
$\det{}' \; \Phi(z,p)$ is independent of the choice of $\{i,j,k\}$ and $\{r,s,t\}$.
One defines a Jacobian factor by
\bq
\label{def_jacobian}
 J\left(z,p\right) & = &
 \frac{1}{\det{}' \; \Phi\left(z,p\right)}.
\eq

\section{The CHY representation of tree-level primitive QCD amplitudes}
\label{sect:CHY_representation_QCD}

We would like to show that all tree-level primitive QCD amplitudes have a representation in the form
\bq
\label{main_result_int_rep}
 A_n\left(w,p,\eps\right)
 & = &
 \frac{i}{\left(2\pi i\right)^{n-3}}
 \int
 \frac{d^nz}{d\omega}
 \;
 \prod{}' \delta\left( f_a\left(z,p\right) \right)
 \;
 \hat{C}\left(w, z\right) 
 \; 
 \hat{E}\left(z,p,\eps\right)
\eq
or equivalently
\bq
\label{main_result}
 A_n\left(w, p, \eps \right)
 & = &
 i
 \sum\limits_{\mathrm{solutions} \; j} J\left(z^{(j)},p\right) \; \hat{C}\left(w, z^{(j)}\right) \; \hat{E}\left(z^{(j)},p,\eps\right).
\eq
In eq.~(\ref{main_result_int_rep}) the symbol $d\omega$ denotes the invariant $\mathrm{PSL}(2,{\mathbb C})$ measure
\bq
 d\omega
 & = &
 \left(-1\right)^{p+q+r}
 \frac{dz_p dz_q dz_r}{\left( z_p - z_q \right) \left( z_q - z_r \right) \left( z_r - z_q \right)}.
\eq
and the primed product of delta functions stands for
\bq
 \prod{}' \delta\left( f_a\left(z,p\right) \right)
 & = & 
 \left(-1\right)^{i+j+k}
 \left( z_i - z_j \right) \left( z_j - z_k \right) \left( z_k - z_i \right)
 \prod\limits_{a \neq i,j,k} \delta\left( f_a\left(z,p\right) \right),
\eq
taking into account that only $(n-3)$ scattering equations are independent.
The form of eq.~(\ref{main_result_int_rep}) or eq.~(\ref{main_result}) can be interpreted as a ``factorisation of information'':
The information on the external polarisations enters only through $\eps$ in $\hat{E}$, 
the information on the external order only through $w$ in $\hat{C}$.
The information on the flavours of the external particles enters $\hat{E}$ (through $\eps$) 
and $\hat{C}$ (through $w$).
The Jacobian $J$ is defined in eq.~(\ref{def_jacobian}).
Under a $\mathrm{PSL}(2,{\mathbb C})$ transformation the Jacobian $J$ transforms
as
\bq
 J\left(g \cdot z, p\right)
 & = &
 \left( \prod\limits_{j=1}^n \frac{1}{\left(c z_j + d\right)^4} \right)
 J\left(z,p\right)
\eq
We require that $\hat{C}$ and $\hat{E}$ transform under $\mathrm{PSL}(2,{\mathbb C})$ transformations as
\bq
\label{PSL2_trafo_C_hat_and_E_hat}
 \hat{C}\left(w, g \cdot z \right)
 & = &
 \left( \prod\limits_{j=1}^n \left(c z_j + d\right)^2 \right)
 \hat{C}\left(w, z\right),
 \nonumber \\
 \hat{E}\left(g \cdot z, p,\eps \right)
 & = &
 \left( \prod\limits_{j=1}^n \left(c z_j + d\right)^2 \right)
 \hat{E}\left(z,p,\eps\right).
\eq
The expression on the right-hand-side of eq.~(\ref{main_result}) is then 
$\mathrm{PSL}(2,{\mathbb C})$ invariant.
We further require that $\hat{E}$ is gauge-invariant.

It will be convenient to introduce the following short-hand notation:
We define a $N_{\mathrm{permutations}}$-dimensional vector $A_w$ with components
\bq
 A_w & = & A_n\left(w, p, \eps \right),
\eq
a $N_{\mathrm{permutations}} \times N_{\mathrm{solutions}}$-dimensional matrix $\hat{M}_{w j}$ by
\bq
 \hat{M}_{w j}
 & = &
 J\left(z^{(j)},p\right) \; \hat{C}\left(w, z^{(j)}\right),
\eq
and a $N_{\mathrm{solutions}}$-dimensional vector $\hat{E}_j$ by
\bq
 \hat{E}_j & = & \hat{E}\left(z^{(j)},p,\eps\right).
\eq
Then eq.~(\ref{main_result}) may be written compactly as
\bq
 A_w & = & i \; \hat{M}_{w j} \hat{E}_j,
\eq
where a sum over $j$ is understood.

\subsection{Normalisation}
\label{sect:normalisation}

We may ask if a representation in the form of eq.~(\ref{main_result}) is unique.
This is certainly not the case. We may always multiply $\hat{C}$ by a non-zero constant $\lambda$
and divide at the same time $\hat{E}$ by the same constant.
More generally, we may multiply $\hat{C}$ by any function of cross-ratios of the variables $z$
and divide $\hat{E}$ by the same function, as long as this function is independent of the 
external ordering.
A function of cross-ratios of the variables $z$ will not change the 
$\mathrm{PSL}(2,{\mathbb C})$ transformation laws in eq.~(\ref{PSL2_trafo_C_hat_and_E_hat}).
In order to eliminate this freedom we make a choice.

We make the choice that the generalised cyclic factor $\hat{C}(w,z)$ agrees with the
standard Parke-Taylor factor $C(w,z)$ for pure gluonic amplitudes ($n_q=0$)
and for amplitudes with one quark-anti-quark-pair ($n_q=1$).
In the massless case these amplitudes are identical to their ${\mathcal N}=4$ SYM counterpart.

For $n_q\ge2$ we make the choice that for amplitudes with the standard orientation of the
fermion lines (defined in eq.~(\ref{def_standard_orientation}))
the generalised cyclic factor $\hat{C}(w,z)$ agrees as well with the standard Parke-Taylor factor $C(w,z)$.
Amplitudes with this orientation of the fermion lines and one fermion line of the highest possible level $(n_q-1)$
are identical to their
single-flavour cousins (any non-trivial permutation of the quarks while keeping the anti-quarks fixed will
lead to crossed fermion lines).
In the massless case these single-flavour amplitudes are in turn identical to their ${\mathcal N}=4$ SYM counterparts
(the couplings of the scalar particles in ${\mathcal N}=4$ are ``flavour-changing'', therefore there is no scalar exchange in single-flavour amplitudes).
(These observations are the basic ideas behind the flavour recursion discussed in \cite{Melia:2013epa}).

\subsection{Definition of $\hat{C}$}
\label{sect:def_C_hat}

In this section we define the generalised cyclic factor $\hat{C}(w,z)$.
We label the external particles of a primitive amplitude $A_n$ by $1$, ..., $n$ and
the associated complex variables $z_j$ occurring in the scattering equations by $z_1$, ..., $z_n$, 
such that the complex variable $z_j$ corresponds to particle $j$.
Our alphabet is then ${\mathbb A} = \{1,2,...,n\}$ 
and a word $w=l_1 l_2 ... l_n \in W_0$ is equivalent to a permutation of $(1,2,...,n)$.
We define the standard cyclic factor $C(w,z)$ for $w=l_1 l_2 ... l_n$
by
\bq
\label{def_standard_C}
 C\left( l_1 l_2 ... l_n, z \right)
 & = &
 \frac{1}{\left( z_{l_1} - z_{l_2} \right) \left( z_{l_2} - z_{l_3} \right) ... \left( z_{l_n} - z_{l_1} \right) }.
\eq
The standard cyclic factor in eq.~(\ref{def_standard_C}) is also called the Parke-Taylor factor.
The standard cyclic factor $C(w,z)$ satisfies for $z$ a solution of the scattering equations
cyclic invariance, the Kleiss-Kuijf relations 
and the fundamental BCJ relations with any choice of letter for $l_2$.
In other words, the standard cyclic factor $C(w,z)$ satisfies all the relations of the pure gluonic primitive tree amplitudes.
The requirement that $z$ is a solution of the scattering equations is needed for the BCJ relations, but not for cyclic
invariance nor for the Kleiss-Kuijf relations.
It will be convenient to view $C(w,z)$ and $\hat{C}(w,z)$ as linear operators on the vector space of words with basis $W_0$,
similar to eq.~(\ref{linear_operator_notation}):
\bq
 C\left(\lambda_1 w_1 + \lambda_2 w_2, z \right)
 & = &
 \lambda_1 C\left(w_1,z\right) + \lambda_2 C\left(w_2,z\right),
 \nonumber \\
 \hat{C}\left(\lambda_1 w_1 + \lambda_2 w_2, z \right)
 & = &
 \lambda_1 \hat{C}\left(w_1,z\right) + \lambda_2 \hat{C}\left(w_2,z\right).
\eq
Let us now give the definition of the generalised cyclic factor $\hat{C}$:
\begin{enumerate}
\item \label{step1}
For $w \in W_3$ we set
\bq
 \hat{C}\left(w,z\right)
 & = &
 C\left(w,z\right),
\eq
i.e. the generalised cyclic factor $\hat{C}(w,z)$ agrees on $W_3$ with the standard cyclic factor $C(w,z)$, in agreement with the comments of section~\ref{sect:normalisation}.
\item \label{step2}
For $w \in W_2 \backslash W_3$ we first define 
\bq
\label{def_C_hat_crossed_fermion_lines}
 \hat{C}\left(w,z\right)
 & = &
 0
\eq
for all words corresponding to crossed fermion lines.
For words with no crossed fermion lines we relate $\hat{C}(w,z)$ as in eq.~(\ref{fermion_orientation}) 
(by substituting $\hat{C}$ for $A_n$)
to a linear combination of $\hat{C}(w_j,z)$'s with $w_j \in W_3$.
With the notation as in section~\ref{sect:Dyck} we have
for sub-words
\bq
 x_k \;\; = \;\; l_{i_1} l_{i_2} ... l_{i_r},
 & &
 y_k \;\; = \;\; l_{j_1} l_{j_2} ... l_{j_s}.
\eq
the relation
\bq
\label{def_C_hat_no_crossed}
\lefteqn{
 \hat{C}\left( x_{k-1} q_i x_k \bar{q}_j w_{k+1} q_j y_k \bar{q}_i y_{k-1}, z \right)
 = } & &
 \nonumber \\
 & &
 \left(-1\right)^{|w_{k+1}|+1}
 \sum\limits_{a=0}^r
 \sum\limits_{b=0}^s
 \hat{C}\left( x_{k-1} q_i l_{i_1} ... l_{i_a} q_j w_{k+1}' \bar{q}_j l_{j_{b+1}} ... l_{j_s} \bar{q}_i y_{k-1}, z \right),
\eq
with
\bq
 w_{k+1}'
 & = &
 \left( l_{i_{a+1}} ... l_{i_r} \right) \shuffle w_{k+1}^T \shuffle \left( l_{j_1} ... l_{j_b} \right).
\eq
This relation allows us to define recursively the generalised cyclic factor for words with $w \in W_2 \backslash W_3$
in terms of generalised cyclic factors of words with $w \in W_3$.
The recursion proceeds along the levels of the fermion lines, as explained in section~\ref{sect:Dyck}.

Eq.~(\ref{def_C_hat_crossed_fermion_lines}) defines $\hat{C}(w,z)$ for words with crossed fermion lines.
For these words $\hat{C}$ is simply zero.
Eq.~(\ref{def_C_hat_no_crossed}) defines recursively the generalised cyclic factor $\hat{C}(w,z)$ for words
with a non-standard orientation of the fermion lines in terms of generalised cyclic factors for words
with a standard orientation of the fermion lines. The latter have already been defined in step \ref{step1}.
\item \label{step3}
For $w \in W_1 \backslash W_2$ we set
\bq
\label{def_C_hat_KK}
 \hat{C}\left(l_1 w_1 l_n w_2, z \right)
 & = &
 \left(-1\right)^{|w_2|} \hat{C}\left( l_1 \left( w_1 \shuffle w_2^T \right) l_n, z \right).
\eq
Eq.~(\ref{def_C_hat_KK}) defines the generalised cyclic factor for words, where the letter $l_n$ does not appear
in the last place in terms of already defined generalised cyclic factors for words, where the letter $l_n$
occurs in the last place.
We recognise in eq.~(\ref{def_C_hat_KK}) the Kleiss-Kuijf relation.
\item \label{step4}
For $w \in W_0 \backslash W_1$ we set
\bq
\label{def_C_hat_cyclic}
 \hat{C}\left(w_1 l_1 w_2, z \right)
 & = &
 \hat{C}\left(l_1 w_2 w_1, z \right).
\eq 
Eq.~(\ref{def_C_hat_cyclic}) defines the generalised cyclic factor for words, where the letter $l_1$ does not appear
in the first place in terms of already defined generalised cyclic factors for words, where the letter $l_1$
occurs in the first place.
We recognise cyclic invariance in eq.~(\ref{def_C_hat_cyclic}).
\end{enumerate}

\subsection{Definition of $\hat{E}$}
\label{sect:def_E_hat}

In this section we define the generalised permutation invariant function $\hat{E}(z,p,\eps)$.
We recall that we defined a $N_{\mathrm{permutations}} \times N_{\mathrm{solutions}}$-dimensional matrix $\hat{M}_{w j}$ by
\bq
 \hat{M}_{w j}
 & = &
 J\left(z^{(j)},p\right) \; \hat{C}\left(w, z^{(j)}\right).
\eq
Let us consider a $N_{\mathrm{basis}} \times N_{\mathrm{solutions}}$-dimensional sub-matrix $\hat{M}^{\mathrm{red}}_{w j}$
by restricting $w \in B$.
Note that we always have
\bq
 N_{\mathrm{basis}} & \le & N_{\mathrm{solutions}},
\eq
therefore the matrix $\hat{M}^{\mathrm{red}}_{w j}$ has less rows than columns.
For $w \in B$ the generalised cyclic factor $\hat{C}$ agrees with the standard Parke-Taylor factor $C$
\bq
 \hat{C}\left(w, z^{(j)}\right)
 & = &
 C\left(w, z^{(j)}\right)
\eq
and the entries of $\hat{M}^{\mathrm{red}}_{w j}$ are given by
\bq
\label{entries_M_w_j}
 \hat{M}^{\mathrm{red}}_{w j}
 & = &
 J\left(z^{(j)},p\right) \; C\left(w, z^{(j)}\right),
 \;\;\;\;\;\; w \in B.
\eq
On a technical level, we will now do the following: 
We first establish that the matrix $\hat{M}^{\mathrm{red}}_{w j}$ has full row rank:
\bq
\label{rank_condition}
 \mathrm{rank} \; \hat{M}^{\mathrm{red}}_{w j} & = & N_{\mathrm{basis}}.
\eq
If $\hat{M}^{\mathrm{red}}_{w j}$ has full row rank, a right-inverse $\hat{N}^{\mathrm{red}}_{j w}$ exists.
The right-inverse might not be unique.
We are interested in a right-inverse $\hat{N}^{\mathrm{red}}_{j w}$ such that the entries in the $j$-th row of
$\hat{N}^{\mathrm{red}}_{j w}$ depend only on $z^{(j)}$, 
but not on the other solutions $z^{(1)}, z^{(2)}, ..., z^{(j-1)}, z^{(j+1)}, ..., z^{(n-3)!}$ of the scattering equations.

The entries of the matrix $\hat{M}^{\mathrm{red}}_{w j}$ are defined in eq.~(\ref{entries_M_w_j})
in terms of the standard Parke-Taylor factor $C(w,z)$ and the Jacobian $J(z,p)$.
Information on the flavour of the particles does not enter the definition 
of the individual entries of the matrix $\hat{M}^{\mathrm{red}}_{w j}$.
The flavour information will only affect the set $B$, giving all the possible first indices $w \in B$ of $\hat{M}^{\mathrm{red}}_{w j}$.
As the flavour information is to a large extent irrelevant, let us for simplicity consider the alphabet
\bq
 {\mathbb A} & = &
 \left\{ 1, 2, ..., n \right\},
\eq
with the implicit understanding that we may recover the information on the flavour of the particles if needed.
The set $W_2$ is then
\bq
 W_2
 & = &
 \left\{ 
  \; l_1 l_2 ... l_n \in W_0 \; | \; l_{1}=1, \; l_n=n \;
 \right\}.
\eq
The set $W_2$ has $(n-2)!$ elements.

Let us first consider the case $n_q \le 2$. For $n_q \le 2$ we have
\bq
 N_{\mathrm{basis}} & = & N_{\mathrm{solutions}},
\eq
and an amplitude basis is given by
\ifthenelse{\boolean{convention_12n}}
{
\bq
 B_{n_q \le 2}
 & = &
 \left\{ 
  \; l_1 l_2 ... l_n \in W_0 \; | \; l_{1}=1, \; l_{2}=2, \; l_n=n \;
 \right\}.
\eq
}
{
\bq
 B_{n_q \le 2}
 & = &
 \left\{ 
  \; l_1 l_2 ... l_n \in W_0 \; | \; l_{1}=1, \; l_{n-1}=n-1, \; l_n=n \;
 \right\}.
\eq
}
The basis $B_{n_q \le 2}$ has $N_{\mathrm{solutions}} = (n-3)!$ elements.
For $n_q \le 2$ the matrix $\hat{M}^{\mathrm{red}}_{w j}$ is a square $N_{\mathrm{solutions}} \times N_{\mathrm{solutions}}$ matrix.
We will need this special matrix in the sequel and we denote this matrix without a hat:
\bq
 M^{\mathrm{red}}_{w j}
 & = &
 J\left(z^{(j)},p\right) \; C\left(w, z^{(j)}\right),
 \;\;\;\;\;\; 
 w \in B_{n_q \le 2}.
\eq
It is known that $M^{\mathrm{red}}$ is invertible. 
We can give an explicit expression for the inverse matrix. 
Let $w = l_1 l_2 ... l_{n-2} l_{n-1} l_n \in B_{n_q \le 2}$ be a word with $l_1=1$, $l_{n-1}=n-1$ and 
$l_n=n$.
We denote by $\bar{w}$ the word
\bq
 \bar{w}
 & = & 
 l_1 l_2 ... l_{n-2} l_n l_{n-1},
\eq
i.e. the word where the last two letters are exchanged.
We then define for $w_1=l_1 ... l_n \in B_{n_q \le 2}$ and $w_2=k_1 ... k_n \in B_{n_q \le 2}$ \cite{Cachazo:2013gna,Kawai:1985xq,BjerrumBohr:2010ta,BjerrumBohr:2010hn}
\bq
 S\left[w_1|\bar{w}_2\right]
 & = &
 \left(-1\right)^n
 \prod\limits_{i=2}^{n-2} 
 \left[
        2 p_{l_1} \cdot p_{l_i} + 2 \Delta_{l_1 l_i}
        + \sum\limits_{j=2}^{i-1} \theta_{\bar{w}_2}\left(l_j,l_i\right) 
                                  \left( 2 p_{l_j} \cdot p_{l_i} + 2 \Delta_{l_j l_i} \right)
 \right],
\eq
with
\bq
\theta_{\bar{w}_2}\left(l_j,l_i\right)
 & = & 
 \left\{
 \begin{array}{ll}
  1 & \mbox{if $l_j$  comes before $l_i$ in the sequence $k_2,k_3,...,k_{n-2}$}, \\
  0 & \mbox{otherwise}. \\
 \end{array}
 \right.
 \nonumber
\eq
We then set 
\bq
\label{def_N_red}
 N^{\mathrm{red}}_{j w}
 & = &
 \sum\limits_{v \in B_{n_q \le 2}}
 S\left[w | \bar{v} \right] C\left( \bar{v}, z^{(j)}\right).
\eq
The $N_{\mathrm{solutions}} \times N_{\mathrm{solutions}}$-dimensional matrix $N^{\mathrm{red}}_{j w}$
is the inverse matrix to $M^{\mathrm{red}}_{w j}$.
Thus we have
\bq
 M^{\mathrm{red}}_{w_1 j} N^{\mathrm{red}}_{j w_2}
 \;\; = \;\;
 \delta_{w_1 w_2},
 & &
 N^{\mathrm{red}}_{j_1 w} M^{\mathrm{red}}_{w j_2}
 \;\; = \;\;
 \delta_{j_1 j_2}.
\eq
Of course, the inverse matrix is unique and a inspection of eq.~(\ref{def_N_red}) shows
that the $j$-th row of $N^{\mathrm{red}}_{j w}$ depends only on $z^{(j)}$ and not on the other solutions
$z^{(i)}$ if $i \neq j$.

Let us now discuss the general case $n_q \in {\mathbb N}_0$.
For $n_q > 2$ we have
\bq
 N_{\mathrm{basis}}
 & < &
 N_{\mathrm{solutions}}
\eq
and the matrix $\hat{M}^{\mathrm{red}}_{w j}$ is now a rectangular 
$N_{\mathrm{basis}} \times N_{\mathrm{solutions}}$-dimensional matrix, with first index given by $w \in B$.
We first have to establish that $\hat{M}^{\mathrm{red}}$ has full row rank, i.e.
\bq
 \mathrm{rank} \; \hat{M}^{\mathrm{red}}_{w j} & = & N_{\mathrm{basis}}.
\eq
This would be easy, if 
\bq
 B & \subseteq & B_{n_q \le 2}.
\eq
However, this is not the case.
For $n_q > 2$ the elements of $B$ do not have a unique letter at 
\ifthenelse{\boolean{convention_12n}}
{position $2$}
{position $(n-1)$}
and in general we have
\bq
 B & \not\subseteq & B_{n_q \le 2}.
\eq
In order to get around this obstruction we recall
that the standard cyclic factors $C(w,z^{(j)})$ satisfy the BCJ relations and we may express the standard
cyclic factor $C(w,z^{(j)})$ for $w \in B$
as a linear combination of standard cyclic factors $C(w',z^{(j)})$ with $w' \in B_{n_q \le 2}$:
\bq
\label{BCJ_relation_cyclic_factor}
 C\left(w,z^{(j)}\right)
 & = &
 F_{w w'} 
 \;
 C\left(w',z^{(j)}\right),
\eq
where a sum over $w' \in B_{n_q \le 2}$ is understood.
$F_{w w'}$ defines a $N_{\mathrm{basis}} \times N_{\mathrm{solutions}}$-dimensional matrix.
The explicit expressions of the entries of $F_{w w'}$ are given in appendix~\ref{sect:def_F_w_wp}.
We note that the entries of the matrix $F_{w w'}$ depend only on the scalar products $2 p_i p_j$, but not on $z^{(j)}$.
We then have
\bq
\label{matrix_splitting}
 \hat{M}^{\mathrm{red}}_{w j}
 & = &
 F_{w w'} M^{\mathrm{red}}_{w' j}.
\eq
The case $n_q \le 2$ is trivially included in eq.~(\ref{matrix_splitting})
by taking $F_{w w'}$ to be the $N_{\mathrm{solutions}} \times N_{\mathrm{solutions}}$ identity matrix.
The matrix $M^{\mathrm{red}}$ has rank $N_{\mathrm{solutions}}$ and is invertible.
It follows that $\hat{M}^{\mathrm{red}}_{w j}$ has rank $N_{\mathrm{basis}}$ if and only if 
the $N_{\mathrm{basis}} \times N_{\mathrm{solutions}}$-matrix $F_{w w'}$ (with $w \in B$ and $w' \in B_{n_q \le 2}$)
has rank $N_{\mathrm{basis}}$.
We have verified for all cases with $n \le 10$ external particles and for generic kinematical configurations that the matrix
$F_{w w'}$ (and hence $\hat{M}^{\mathrm{red}}_{w j}$) has rank $N_{\mathrm{basis}}$.
Based on this evidence we will in the sequel assume that 
$F_{w w'}$ has rank $N_{\mathrm{basis}}$:
\bq
\label{rank_condition_F}
 \mathrm{rank} \; F_{w w'} & = & N_{\mathrm{basis}},
\eq
Note that eq.~(\ref{rank_condition_F}) is a purely kinematical statement, independent of flavour
and independent of the variables $z^{(j)}$.
We further note that by a suitable ordering of the bases $B$ and $B_{n_q \le 2}$ the
matrix $F_{w w'}$ can be brought into an upper triangle block structure.
It is therefore sufficient to show that all (square) matrices on the main diagonal have full rank.
The details are given in appendix~\ref{sect:upper_triangle_block}.

Assuming from now on that the matrix $F_{w w'}$ has maximal row rank, the 
$N_{\mathrm{basis}} \times N_{\mathrm{basis}}$-dimensional matrix $F F^T$ is invertible and
the $N_{\mathrm{solutions}} \times N_{\mathrm{basis}}$-dimensional matrix
\bq
\label{def_G}
 G & = & F^T \left( F F^T \right)^{-1}
\eq
defines a right inverse to $F$:
\bq
 F_{w_1 w'} G_{w' w_2}
 & = &
 \delta_{w_1 w_2}.
\eq
We then set
\bq
\label{def_N_hat_red}
 \hat{N}^{\mathrm{red}}
 & = &
 N^{\mathrm{red}} G.
\eq
The $N_{\mathrm{solutions}} \times N_{\mathrm{basis}}$-dimensional matrix $\hat{N}^{\mathrm{red}}$ is then a
right inverse to $\hat{M}^{\mathrm{red}}$:
\bq
\label{matrix_identity}
 \hat{M}^{\mathrm{red}}_{w_1 j} \hat{N}^{\mathrm{red}}_{j w_2}
 & = &
 \delta_{w_1 w_2}.
\eq
Having defined $\hat{N}^{\mathrm{red}}_{j w}$, we set
\bq
\label{definition_E_hat_j}
 \hat{E}_j 
 & = &
 - i \hat{N}^{\mathrm{red}}_{j w} A_w,
\eq
where a sum over all $w \in B$ is understood.
Putting everything together, we arrive along the lines of ref.~\cite{Weinzierl:2014ava} at the definition of
the generalised permutation invariant function $\hat{E}(z,p,\eps)$:
\bq
\label{definition_E_hat}
 \hat{E}\left(z,p,\eps\right)
 & = &
 - i
 \;\;
 \sum\limits_{u,v \in B_{n_q \le 2}}
 \;\;
 \sum\limits_{w \in B}
 \;\;
 S\left[ u | \bar{v} \right] G_{u w} C\left( \bar{v}, z \right) A_n\left(w,p,\eps\right).
\eq
A few comments are in order:
The attentive reader may ask, why we did not simply define $\hat{N}^{\mathrm{red}}$
as
\bq
\label{naive_right_inverse}
 \hat{M}^{\mathrm{red}}{}^T \left( \hat{M}^{\mathrm{red}} \hat{M}^{\mathrm{red}}{}^T \right)^{-1}.
\eq
The reason is as follows: We would like to have that $\hat{E}_j$ depends only on the $j$-th solution
of the scattering equations, but not on all the other solutions.
Within our definition this is manifest.
$F$ and $G$ are independent of $z$, and so is $S[w_1|\bar{w}_2]$.
The $z$-dependence comes entirely from $C\left( \bar{v}, z^{(j)}\right)$ in eq.~(\ref{def_N_red}).
Therefore $\hat{N}^{\mathrm{red}}_{j w}$ depends only on $z^{(j)}$ and not on $z^{(i)}$ if $ i \neq j$.
We can therefore define a function $\hat{E}\left(z,p,\eps\right)$ on $\hat{\mathbb C}^n$ 
as done in eq.~(\ref{definition_E_hat}).
On the other hand, this is far from clear for the expression in eq.~(\ref{naive_right_inverse}).

A second comment is related to the uniqueness of our definition in eq.~(\ref{definition_E_hat}).
For $n_q > 2$ the right-inverse $G_{w' w}$ to the matrix $F_{w w'}$ is not unique.
It is of course unique for invertible matrices, i.e. the case $n_q \le 2$.
We may parametrise the general form of the right-inverse as
\bq
 G_{w' w} + \left( \delta_{w' w_2'} - G_{w' w_1} F_{w_1 w_2'} \right) X_{w_2' w}
\eq
with an arbitrary $N_{\mathrm{solutions}} \times N_{\mathrm{basis}}$-dimensional matrix $X_{w' w}$.
Plugging this into eq.~(\ref{definition_E_hat}) we find
\bq
 \hat{E}\left(z,p,\eps\right)
 & \rightarrow &
 \hat{E}\left(z,p,\eps\right)
 - i
 \;\;
 \sum\limits_{u,v \in B_{n_q \le 2}}
 \;\;
 S\left[ u | \bar{v} \right] 
 \left( \delta_{u w_2'} - G_{u w_1} F_{w_1 w_2'} \right) x_{w_2'}
 C\left( \bar{v}, z \right),
\eq
or equivalently
\bq
 \hat{E}_j
 & \rightarrow &
 \hat{E}_j
 -
 i N^{\mathrm{red}}_{j w'} \left( \delta_{w' w_2'} - G_{w' w_1} F_{w_1 w_2'} \right) x_{w_2'},
\eq
with some arbitrary $N_{\mathrm{solutions}}$-dimensional vector $x_{w'}$.
This arbitrariness does not affect expressions of the form
\bq
\label{expr_with_Y}
 i
 \sum\limits_{\mathrm{solutions} \; j} J\left(z^{(j)},p\right) \; \hat{Y}\left(z^{(j)}\right) \; \hat{E}\left(z^{(j)},p,\eps\right),
\eq
as long as $\hat{Y}$ has an expansion in $\hat{C}(w,z^{(j)})$ with $w \in B$:
\bq
\label{cond_Y}
 \hat{Y}\left(z^{(j)}\right)
 & = &
 \sum\limits_{w \in B} c_w \hat{C}\left(w,z^{(j)}\right).
\eq
Then we may write
\bq
 J\left(z^{(j)},p\right) \; \hat{Y}\left(z^{(j)}\right)
 & = &
 \sum\limits_{w \in B} c_w \hat{M}^{\mathrm{red}}_{w j}
\eq
and we have
\bq
 i
 \sum\limits_{\mathrm{solutions} \; j}
 \sum\limits_{w \in B}
 c_w \hat{M}^{\mathrm{red}}_{w j}
 \left[
  \hat{E}_j
  -
  i N^{\mathrm{red}}_{j w'} \left( \delta_{w' w_2'} - G_{w' w_1} F_{w_1 w_2'} \right) x_{w_2'}
 \right]
 & = &
 i
 \sum\limits_{\mathrm{solutions} \; j}
 \sum\limits_{w \in B}
 c_w \hat{M}^{\mathrm{red}}_{w j}
  \hat{E}_j,
 \nonumber \\
\eq
since
\bq
 \hat{M}^{\mathrm{red}}_{w j} N^{\mathrm{red}}_{j w'}
 \;\; = \;\;
 F_{w w_3'} M^{\mathrm{red}}_{w_3' j} N^{\mathrm{red}}_{j w'}
 \;\; = \;\;
 F_{w w'}
 & \mbox{and} &
 F_{w w'} \left( \delta_{w' w_2'} - G_{w' w_1} F_{w_1 w_2'} \right) 
 \;\; = \;\;
 0.
\eq
For the tree-level primitive QCD amplitudes we will always have that the factor $\hat{Y}$ appearing in the
sum as in eq.~(\ref{expr_with_Y}) is of the form as in eq.~(\ref{cond_Y}) with 
$w \in B$ for $\hat{C}(w,z^{(j)})$. Therefore the non-uniqueness of the right-inverse does not affect
tree-level primitive QCD amplitudes.

\subsection{Proof of the CHY representation}
\label{sect:proof_CHY_representation}

Let us set
\bq
\label{def_A_tilde}
 \tilde{A}_n\left(w\right)
 & = &
 i
 \sum\limits_{\mathrm{solutions} \; j} J\left(z^{(j)},p\right) \; \hat{C}\left(w, z^{(j)}\right) \; \hat{E}\left(z^{(j)},p,\eps\right),
\eq
with $\hat{C}$ defined in section~\ref{sect:def_C_hat} and $\hat{E}$ defined in section~\ref{sect:def_E_hat}.
We would like to show that
\bq
\label{to_be_shown}
 \tilde{A}_n\left(w\right)
 & = &
 A_n\left(w\right),
 \;\;\;\;\;\;
 \forall w \in W_0.
\eq
It is sufficient to check the five conditions at the end of section~\ref{sect:amplitude_basis}.
\begin{enumerate}
\item We start with $w \in B$. 
We have
\bq
 \tilde{A}_n\left(w\right)
 \;\; = \;\;
 i \hat{M}_{w j} \hat{E}_j 
 \;\; = \;\;
 \hat{M}_{w j} \hat{N}^{\mathrm{red}}_{j w'} A_{w'}.
\eq
Since $w \in B$ we may replace the matrix row $\hat{M}_{w j}$ with the matrix row
$\hat{M}^{\mathrm{red}}_{w j}$ (the two rows are identical).
We therefore have
\bq
 \tilde{A}_n\left(w\right)
 \;\; = \;\;
 \hat{M}^{\mathrm{red}}_{w j} \hat{N}^{\mathrm{red}}_{j w'} A_{w'}
 \;\; = \;\;
 A_{w},
\eq
where we used eq.~(\ref{matrix_identity}). Switching back to the word notation we have
\bq
 \tilde{A}_n\left(w\right)
 & = &
 A_n\left(w\right).
\eq
\item Let us now consider $w \in W_3 \backslash B$.
We have to verify the fundamental BCJ relation:
\bq
 \sum\limits_{i=2}^{n-1} 
  \left( \sum\limits_{k=i+1}^n 2 p_2 p_k \right)
  \tilde{A}_n\left( l_1 l_3 ... l_i l_2 l_{i+1} ... l_{n-1} l_n \right)
 & = & 0.
\eq
In the definition of $\tilde{A}_n$ only $\hat{C}$ depends on the cyclic order and therefore we should have
\bq
 \sum\limits_{i=2}^{n-1} 
  \left( \sum\limits_{k=i+1}^n 2 p_2 p_k \right)
  \hat{C}\left( l_1 l_3 ... l_i l_2 l_{i+1} ... l_{n-1} l_n, z^{(j)} \right)
 & = & 0
\eq
for all solutions $z^{(j)}$ of the scattering equations.
For $w \in W_3$ the cyclic factor $\hat{C}$ agrees with the standard Parke-Taylor factor:
\bq
 \hat{C}\left(w,z\right)
 & = &
 C\left(w,z\right).
\eq
The validity of 
\bq
 \sum\limits_{i=2}^{n-1} 
  \left( \sum\limits_{k=i+1}^n 2 p_2 p_k \right)
  C\left( l_1 l_3 ... l_i l_2 l_{i+1} ... l_{n-1} l_n, z^{(j)} \right)
 & = & 0
\eq
can be inferred from the pure gluon case.
Note that we have to require that the $z^{(j)}$'s are solutions of the scattering equations.
\item Let us now consider $w \in W_2 \backslash W_3$.
We have defined $\hat{C}(w,z)=0$ whenever $w$ corresponds to an external ordering with
crossed fermion lines.
This implies 
\bq
 \tilde{A}_n\left(w \right) & = & 0
\eq
for words corresponding to crossed fermion lines. For words $w \in W_2 \backslash W_3$ 
with no crossed fermion lines
we have defined $\hat{C}$ through eq.~(\ref{def_C_hat_no_crossed}).
As $\tilde{A}_n(w)$ depends on the external ordering only through $\hat{C}(w,z)$, a similar relation
holds for $\tilde{A}_n(w)$.
In other words, $\tilde{A}_n$ satisfies eq.~(\ref{fermion_orientation}).
\item We may repeat this argumentation for $w \in W_1 \backslash W_2$ and afterwards for $w \in W_0 \backslash W_1$.
In both cases we have defined $\hat{C}(w,z)$ such that the required relations (Kleiss-Kuijf relations for $w \in W_1 \backslash W_2$ and cyclic invariance for $w \in W_0 \backslash W_1$) are fulfilled.
\end{enumerate}
This completes the proof of eq.~(\ref{to_be_shown}) and we have shown that any tree-level
primitive QCD amplitude has a CHY representation in the form of eq.~(\ref{main_result}),
with $\hat{C}$ defined in section~\ref{sect:def_C_hat}
and $\hat{E}$ defined in section~\ref{sect:def_E_hat}.

The generalised cyclic factor $\hat{C}(w,z)$ defined in section~\ref{sect:def_C_hat} is always a linear combination
of standard Parke-Taylor factors $C(w,z)$ with $z$-independent coefficients.
Since the standard Parke-Taylor factors $C(w,z)$ transform under $\mathrm{PSL}(2,{\mathbb C})$ transformations
as in eq.~(\ref{PSL2_trafo_C_hat_and_E_hat}), it follows that $\hat{C}(w,z)$ transforms as well as in
eq.~(\ref{PSL2_trafo_C_hat_and_E_hat}).
A similar argument applies to the $\mathrm{PSL}(2,{\mathbb C})$ transformation properties of $\hat{E}(z,p,\eps)$.
Eq.~(\ref{definition_E_hat}) shows that $\hat{E}(z,p,\eps)$ is a linear combination
of standard Parke-Taylor factors $C(\bar{v},z)$ with $z$-independent coefficients.
Therefore it follows that $\hat{E}(z,p,\eps)$ transforms as in eq.~(\ref{PSL2_trafo_C_hat_and_E_hat})
under $\mathrm{PSL}(2,{\mathbb C})$ transformations.

Finally, let us comment on the gauge invariance of $\hat{E}(z,p,\eps)$:
In section~\ref{sect:def_E_hat} we defined $\hat{E}(w,p,\eps)$ in terms of amplitudes $A_n(w)$ from the basis $w \in B$.
The amplitudes are gauge-invariant and the gauge-invariance of $\hat{E}(z,p,\eps)$ follows trivially.
 
\section{An example}
\label{sect:example}

We would like to illustrate our construction with a concrete example.
A non-trivial example is the six-point amplitude $A_6$ with three quark-anti-quark-pairs.
We label the external particles from $1$ to $6$ and we set
\bq
 q_1 \; = \; 1,
 \;\;\;
 q_2 \; = \; 2,
 \;\;\;
 q_3 \; = \; 3,
 \;\;\;
 \bar{q}_3 \; = \; 4,
 \;\;\;
 \bar{q}_2 \; = \; 5,
 \;\;\;
 \bar{q}_1 \; = \; 6.
\eq
Our alphabet is then
\bq
 {\mathcal A}
 \;\; = \;\;
 \left\{ q_1, q_2, q_3, \bar{q}_3, \bar{q}_2, \bar{q}_1 \right\}
 \;\; = \;\;
 \left\{ 1,2,3,4,5,6 \right\}.
\eq
The basis $B$ consists of four elements:
\bq
 B & = &
 \left\{ 
  1 2 3 4 5 6,
  1 2 5 3 4 6,
  1 3 2 5 4 6,
  1 3 4 2 5 6
 \right\}.
\eq
The set $B_{n_q \le 2}$ contains six elements:
\bq
 B_{n_q \le 2}
 & = &
 \left\{
  1 2 3 4 5 6,
  1 2 4 3 5 6,
  1 3 2 4 5 6,
  1 3 4 2 5 6,
  1 4 2 3 5 6,
  1 4 3 2 5 6
 \right\}.
\eq
Note that in the basis $B$ we will have either particle $4$ or particle $5$ at position $5$, while all elements in the set $B_{n_q \le 2}$
have particle $5$ at position $5$.
Since the permutation invariant function $\hat{E}(z,p,\eps)$ involves Parke-Taylor factors with particle $6$ at position $5$ and
particle $5$ at position $6$ we introduce the set $\bar{B}_{n_q \le 2}$ given by
\bq
 \bar{B}_{n_q \le 2}
 & = &
 \left\{
  1 2 3 4 6 5,
  1 2 4 3 6 5,
  1 3 2 4 6 5,
  1 3 4 2 6 5,
  1 4 2 3 6 5,
  1 4 3 2 6 5
 \right\}.
\eq
The set $\bar{B}_{n_q \le 2}$ is just the set $B_{n_q \le 2}$ with particles $5$ and $6$ exchanged.
The permutation invariant function is then given as a double sum in amplitudes $A_n(w,p,\eps)$ from the basis $w \in B$
and Parke-Taylor factors $C(\bar{v},z)$ from the set $\bar{v} \in \bar{B}_{n_q \le 2}$ as
\bq
 \hat{E}\left(z,p,\eps\right)
 & = &
 - i
 \sum\limits_{\bar{v} \in \bar{B}_{n_q \le 2}}
 \sum\limits_{w \in B}
 c_{\bar{v} w}\left(p\right)
 C\left(\bar{v},z\right)
 A_n\left(w,p,\eps\right).
\eq
The coefficients $c_{\bar{v} w}(p)$ depend only on the kinematical variables $2 p_i p_j$ (and the masses $m_j$)
and are given by
\bq
 c_{\bar{v} w}\left(p\right)
 & = &
 \sum\limits_{u \in B_{n_q \le 2}}
 S\left[ u | \bar{v} \right] G_{u w}.
\eq
Due to the inverse matrix in eq.~(\ref{def_G}) the explicit expressions for $c_{\bar{v} w}(p)$
are rather long and not reported here.

Let us now consider the generalised cyclic factor $\hat{C}(w,z)$. For $w \in B$ the generalised cyclic factor agrees
with the standard Parke-Taylor factor. If $w$ corresponds to an external ordering with crossed fermion lines, the generalised cyclic factor
equals zero.
Let us therefore consider as an example the word $w=153426$. This word does not correspond to crossed fermion lines.
However the fermion line $2$-$5$ does not have the standard orientation.
With the definitions of section~\ref{sect:def_C_hat} we have
\bq
 \hat{C}\left(153426,z\right)
 \;\; = \;\;
 - \hat{C}\left(124356,z\right)
 \;\; = \;\;
 \hat{C}\left(123456,z\right)
 \;\; = \;\;
 C\left(123456,z\right).
\eq

\section{Conclusions}
\label{sect:conclusions}

In this paper we have shown that a CHY representation exists for all tree-level primitive QCD amplitudes.
We provided a definition of the generalised cyclic factor $\hat{C}(w,z)$ 
and a definition of the generalised permutation invariant function $\hat{E}(z,p,\eps)$.
The  virtue of the CHY representation 
lies in the fact that it separates the information on the external ordering (contained in the generalised cyclic factor $\hat{C}(w,z)$)
from the information on the helicities of the external particles (contained in the generalised permutation invariant function 
$\hat{E}(z,p,\eps)$).

\subsection{Acknowledgements}

L.d.l.C. is grateful for financial support from CONACYT and the DAAD.

\begin{appendix}

\section{Orientation of fermion lines}
\label{sect:fermion_orientation}

In this appendix we prove eq.~(\ref{fermion_orientation}).
A slightly modified form of eq.~(\ref{fermion_orientation}) has been stated in \cite{Melia:2013epa} and the idea
of the proof can be found in \cite{Melia:2013bta}.
We consider
\bq
 A_n & = &
 A_n\left( x_{k-1} q_i x_k \bar{q}_j w_{k+1} q_j y_k \bar{q}_i y_{k-1} \right).
\eq
Let us assume that the sub-words $x_k$ and $y_k$ consist of $r$ letters and $s$ letters, respectively:
\bq
 x_k \;\; = \;\; l_{i_1} l_{i_2} ... l_{i_r},
 & &
 y_k \;\; = \;\; l_{j_1} l_{j_2} ... l_{j_s}.
\eq
It will be convenient to set
\bq
 w_{k-1} & = & y_{k-1} x_{k-1}.
\eq
Using cyclic invariance we have
\bq
 A_n
 & = &
 A_n\left( q_i x_k \bar{q}_j w_{k+1} q_j y_k \bar{q}_i w_{k-1} \right).
\eq
We now use the Kleiss-Kuijf relation to flip $x_k$:
\bq
 A_n
 & = &
 \left(-1\right)^r
 A_n\left( q_i \bar{q}_j \left( w_{k+1} q_j y_k \bar{q}_i w_{k-1} \right) \shuffle \left( l_{i_r} ... l_{i_1} \right)  \right).
\eq
If we would work out the shuffle product, we would obtain words, where the first $a$ letters of $x_k$ occur after $q_j$
and the remaining $(r-a)$ letters of $x_k$ occur before $q_j$, with $a$ ranging from $0$ to $r$.
Writing this out we have
\bq
 A_n
 & = &
 \left(-1\right)^r
 \sum\limits_{a=0}^r
 A_n\left( 
          q_i \bar{q}_j 
          \left[ w_{k+1} \shuffle \left( l_{i_r} ... l_{i_{a+1}} \right) \right]
          q_j 
          \left[ \left(  y_k \bar{q}_i w_{k-1} \right) \shuffle \left( l_{i_a} ... l_{i_1} \right) \right]
 \right).
\eq
We then use a second time the Kleiss-Kuijf relation to flip the 
sub-word $\bar{q}_j [ w_{k+1} \shuffle ( l_{i_r} ... l_{i_{a+1}} ) ]$:
\bq
 A_n
 =
 \sum\limits_{a=0}^r
 \left(-1\right)^{|w_{k+1}|+1-a}
 A_n\left( 
          q_i 
          q_j 
          \left\{
          \left[ w_{k+1}^T \shuffle \left( l_{i_{a+1}} ... l_{i_r} \right) \right]
          \bar{q}_j 
          \right\}
          \shuffle
          \left[ \left(  y_k \bar{q}_i w_{k-1} \right) \shuffle \left( l_{i_a} ... l_{i_1} \right) \right]
 \right).
\eq
The shuffle product is associative and therefore
\bq
 A_n
 =
 \sum\limits_{a=0}^r
 \left(-1\right)^{|w_{k+1}|+1-a}
 A_n\left( 
          q_i 
          q_j 
          \left\{
          \left[ w_{k+1}^T \shuffle \left( l_{i_{a+1}} ... l_{i_r} \right) \right]
          \bar{q}_j 
          \right\}
          \shuffle
          \left(  y_k \bar{q}_i w_{k-1} \right) \shuffle \left( l_{i_a} ... l_{i_1} \right) 
 \right).
\eq
We may then use the (inverse) Kleiss-Kuijf relation to bring back $(l_{i_a} ... l_{i_1})$ between $q_i$ and $q_j$:
\bq
 A_n
 & = &
 \left(-1\right)^{|w_{k+1}|+1}
 \sum\limits_{a=0}^r
 A_n\left( 
          q_i 
          l_{i_1} ... l_{i_a}  
          q_j 
          \left\{
          \left[ w_{k+1}^T \shuffle \left( l_{i_{a+1}} ... l_{i_r} \right) \right]
          \bar{q}_j 
          \right\}
          \shuffle
          \left(  y_k \bar{q}_i w_{k-1} \right)
 \right).
\eq
In the shuffle product of $[ w_{k+1}^T \shuffle ( l_{i_{a+1}} ... l_{i_r} ) ] \bar{q}_j$ with $y_k \bar{q}_i w_{k-1}$ 
only the terms where $\bar{q}_j$ occurs before $\bar{q}_i$ are non-zero.
The other terms have a crossed fermion line and the amplitude is zero for those.
Writing the sub-word $y_k$ in terms of letters we obtain
\bq
\lefteqn{
 A_n
 = 
} & & \\
 & &
 \left(-1\right)^{|w_{k+1}|+1}
 \sum\limits_{a=0}^r
 \sum\limits_{b=0}^s
 A_n\left( 
          q_i 
          l_{i_1} ... l_{i_a}  
          q_j 
          \left[
           \left( l_{i_{a+1}} ... l_{i_r} \right) 
           \shuffle
           w_{k+1}^T
           \shuffle
           \left( l_{j_1} ... l_{j_b} \right)
          \right]
          \bar{q}_j 
          l_{j_{b+1}} ... l_{j_s} 
          \bar{q}_i w_{k-1}
 \right).
 \nonumber
\eq
Finally, using cyclic invariance one arrives at
\bq
 A_n & = &
 \left(-1\right)^{|w_{k+1}|+1}
 \sum\limits_{a=0}^r
 \sum\limits_{b=0}^s
 A_n\left( x_{k-1} q_i l_{i_1} ... l_{i_a} q_j w_{k+1}' \bar{q}_j l_{j_{b+1}} ... l_{j_s} \bar{q}_i y_{k-1} \right),
\eq
with
\bq
 w_{k+1}'
 & = &
 \left( l_{i_{a+1}} ... l_{i_r} \right) \shuffle w_{k+1}^T \shuffle \left( l_{j_1} ... l_{j_b} \right).
\eq

\section{The matrix $F_{w w'}$}
\label{sect:def_F_w_wp}

\ifthenelse{\boolean{convention_12n}}
{
In this appendix we define the entries of the matrix $F_{w w'}$, occurring in eq.~(\ref{BCJ_relation_cyclic_factor}).
We may neglect flavour and it is therefore convenient to consider the alphabet
\bq
 {\mathbb A} 
 & = &
 \left\{ 1, 2, ..., n \right\}.
\eq
We set as before
\bq
 W_0
 & = &
 \left\{ \; l_1 l_2 ... l_n \; | \; l_i \in {\mathbb A}, \; l_i \neq l_j \;\mbox{for} \; i \neq j \; \right\}
\eq
and
\bq
 W_2
 & = &
 \left\{ 
  \; l_1 l_2 ... l_n \in W_0 \; | \; l_{1}=1, \; l_n=n \;
 \right\},
 \nonumber \\
 B 
 & = &
 \left\{ 
  \; l_1 l_2 ... l_n \in W_0 \; | \; l_{1}=1, \; l_{2}=2, \; l_n=n \;
 \right\}.
\eq
For a sub-word $w=l_1 l_2 ... l_k$ we set
\bq
 S\left(w\right)
 & = &
 \sum\limits_{\sigma \in S_k}
 l_{\sigma(1)} l_{\sigma(2)} ... l_{\sigma(k)}.
\eq
Let $w_1=l_1 l_2 ... l_j$ and $w_2=l_{j+1} l_{j+2} ... l_{n-3}$ be two sub-words, such that $w = 1 w_1 2 w_2 n \in W_2$.
For convenience we set $l_0=2$.
The standard cyclic factors $C(w,z^{(j)})$ satisfy the BCJ relations and we have
\bq
\label{BCJ_relation_C}
 C\left( w, z^{(j)} \right)
 & = &
 \sum\limits_{w'} F_{w w'}
 C\left( w', z^{(j)} \right).
\eq
The sum is over all words occurring in
\bq
 1 2 \left( S(w_1) \shuffle w_2 \right) n. 
\eq
For a given $w$ we define $F_{w w'}=0$ if $w'$ does not appear in the sum of eq.~(\ref{BCJ_relation_C}).   
Otherwise, the coefficients are given for $w'= 1 2 \sigma_1 \sigma_2 ... \sigma_{n-3} n = 1 2 \sigma n$ by \cite{Bern:2008qj}
\bq
 F_{w w'}
 & = &
 \prod\limits_{k=1}^j \frac{{\mathcal F}\left(2 \sigma n | l_k\right)}{\hat{s}_{1, l_1,...,l_k}},
\eq
where for $\rho = 2 \sigma n$ the function ${\mathcal F}(\rho | l_k)$ is given by
\bq
\lefteqn{
 {\mathcal F}\left( \rho | l_k \right)
 = } & &
 \\
 & &
 \left\{
  \begin{array}{rl}
     \sum\limits_{r=t_{l_k}+1}^{n-1} {\mathcal G}\left( l_k, \rho_r \right) & \mbox{if} \; t_{l_{k-1}} < t_{l_k} \\
   - \sum\limits_{r=1}^{t_{l_k}-1} {\mathcal G}\left( l_k, \rho_r \right) & \mbox{if} \; t_{l_{k-1}} > t_{l_k} \\
  \end{array}
 \right\}
 + 
 \left\{
  \begin{array}{rl}
     \hat{s}_{1, l_1,...,l_k} & \mbox{if} \; t_{l_{k-1}} < t_{l_k} < t_{l_{k+1}} \\
   - \hat{s}_{1, l_1,...,l_k} & \mbox{if} \; t_{l_{k-1}} > t_{l_k} > t_{l_{k+1}} \\
                 0 & \mbox{else}
  \end{array}
 \right\}.
 \nonumber 
\eq
$t_{a}$ denotes the position of leg $a$ in the string $\rho$, except for $t_{l_0}$ and $t_{l_{j+1}}$, which are always
defined to be
\bq
 t_{l_0} \;\; = \;\; t_{l_2},
 & &
 t_{l_{j+1}} \;\; = \;\; 0.
\eq
For $j=1$ this implies
\bq
 t_{l_0}
 \;\; = \;\;
 t_{l_2}
 \;\; = \;\;
 0.
\eq
The function ${\mathcal G}$ is given by
\bq
 {\mathcal G}\left( l_k, \rho_r \right)
 & = &
 \left\{
  \begin{array}{rl}
   2 p_{l_k} p_{\rho_r} + 2 \Delta_{l_k \rho_r} & \mbox{if} \; \rho_r = 2,n \\
   2 p_{l_k} p_{\rho_r} + 2 \Delta_{l_k \rho_r} & \mbox{if} \; \rho_r = l_t \; \mbox{and} \; k < t \\
   0 & \mbox{else}
 \end{array}
 \right\}.
\eq
We used the notation
\bq
 \hat{s}_{\alpha_1,...,\alpha_k}
 & = &
 \sum\limits_{i<j}
 \left( 2 p_{\alpha_i} p_{\alpha_j} + 2 \Delta_{\alpha_i \alpha_j} \right).
\eq
Let us mention that the coefficients $F_{w w'}$ are the ones appearing in the general BCJ relations for tree-level primitive
QCD amplitudes \cite{Bern:2008qj,Johansson:2015oia}.
We presented them here in a form which holds also for the massive case.
The general form of the BCJ relations is as follows:
Let $w_1=l_1 l_2 ... l_j$ be a sub-word consisting only of gluon legs 
and $w_2=l_{j+1} l_{j+2} ... l_{n-3}$ a second sub-word, where particles of any type may occur.
We further assume that $w = 1 w_1 2 w_2 n \in W_2$.
The general BCJ relation reads
\bq
\label{general_BCJ}
 A_n\left( w \right)
 & = &
 \sum\limits_{w'} F_{w w'}
 A_n\left( w' \right).
\eq
As before, the sum is over all words occurring in
\bq
 1 2 \left( S(w_1) \shuffle w_2 \right) n,
\eq
and the coefficients $F_{w w'}$ are defined as above.
The general BCJ relations of eq.~(\ref{general_BCJ}) follow from the fundamental BCJ relations \cite{Feng:2010my}.
}
{
In this appendix we define the entries of the matrix $F_{w w'}$, occurring in eq.~(\ref{BCJ_relation_cyclic_factor}).
We may neglect flavour and it is therefore convenient to consider the alphabet
\bq
 {\mathbb A} 
 & = &
 \left\{ 1, 2, ..., n \right\}.
\eq
We set as before
\bq
 W_0
 & = &
 \left\{ \; l_1 l_2 ... l_n \; | \; l_i \in {\mathbb A}, \; l_i \neq l_j \;\mbox{for} \; i \neq j \; \right\}
\eq
and
\bq
 W_2
 & = &
 \left\{ 
  \; l_1 l_2 ... l_n \in W_0 \; | \; l_{1}=1, \; l_n=n \;
 \right\},
 \nonumber \\
 B 
 & = &
 \left\{ 
  \; l_1 l_2 ... l_n \in W_0 \; | \; l_{1}=1, \; l_{n-1}=n-1, \; l_n=n \;
 \right\}.
\eq
For a sub-word $w=l_1 l_2 ... l_k$ we set
\bq
 S\left(w\right)
 & = &
 \sum\limits_{\sigma \in S_k}
 l_{\sigma(1)} l_{\sigma(2)} ... l_{\sigma(k)}.
\eq
Let $w_1=l_1 l_2 ... l_j$ and $w_2=l_{j+1} l_{j+2} ... l_{n-3}$ be two sub-words, such that $w = 1 w_1 (n-1) w_2 n \in W_2$.
For convenience we set $l_{n-2}=n-1$.
The standard cyclic factors $C(w,z^{(j)})$ satisfy the BCJ relations and we have
\bq
\label{BCJ_relation_C}
 C\left( w, z^{(j)} \right)
 & = &
 \sum\limits_{w'} F_{w w'}
 C\left( w', z^{(j)} \right).
\eq
The sum is over all words occurring in
\bq
 1 \left( w_1 \shuffle S(w_2) \right) (n-1) n. 
\eq
For a given $w$ we define $F_{w w'}=0$ if $w'$ does not appear in the sum of eq.~(\ref{BCJ_relation_C}).   
Otherwise, the coefficients are given for $w'= 1 \sigma_1 \sigma_2 ... \sigma_{n-3} (n-1) n = 1 \sigma (n-1) n$ by \cite{Bern:2008qj}
\bq
 F_{w w'}
 & = &
 \prod\limits_{k=j+1}^{n-3} \frac{{\mathcal F}\left(1 \sigma (n-1) | l_k\right)}{\hat{s}_{n,l_k,...,l_{n-3}}},
\eq
where for $\rho = 1 \sigma (n-1)$ the function ${\mathcal F}(\rho | l_k)$ is given by
\bq
\lefteqn{
 {\mathcal F}\left( \rho | l_k \right)
 = } & &
 \\
 & &
 \left\{
  \begin{array}{rl}
     \sum\limits_{r=1}^{t_{l_k}-1} {\mathcal G}\left( l_k, \rho_r \right) & \mbox{if} \; t_{l_{k}} < t_{l_{k+1}} \\
   - \sum\limits_{r=t_{l_k}+1}^{n-1} {\mathcal G}\left( l_k, \rho_r \right) & \mbox{if} \; t_{l_{k}} > t_{l_{k+1}} \\
  \end{array}
 \right\}
 + 
 \left\{
  \begin{array}{rl}
     \hat{s}_{n,l_k,...,l_{n-3}} & \mbox{if} \; t_{l_{k-1}} < t_{l_k} < t_{l_{k+1}} \\
   - \hat{s}_{n,l_k,...,l_{n-3}} & \mbox{if} \; t_{l_{k-1}} > t_{l_k} > t_{l_{k+1}} \\
                 0 & \mbox{else}
  \end{array}
 \right\}.
 \nonumber 
\eq
$t_{a}$ denotes the position of leg $a$ in the string $\rho$, except for $t_{l_{n-2}}$ and $t_{l_j}$, which are always
defined to be
\bq
 t_{l_{n-2}} \;\; = \;\; t_{l_{n-4}},
 & &
 t_{l_j} \;\; = \;\; n.
\eq
For $j=n-4$ this implies
\bq
 t_{l_{n-2}}
 \;\; = \;\;
 t_{l_{n-4}}
 \;\; = \;\;
 n.
\eq
The function ${\mathcal G}$ is given by
\bq
 {\mathcal G}\left( l_k, \rho_r \right)
 & = &
 \left\{
  \begin{array}{rl}
   2 p_{l_k} p_{\rho_r} + 2 \Delta_{l_k \rho_r} & \mbox{if} \; \rho_r = 1,(n-1) \\
   2 p_{l_k} p_{\rho_r} + 2 \Delta_{l_k \rho_r} & \mbox{if} \; \rho_r = l_t \; \mbox{and} \; t < k \\
   0 & \mbox{else}
 \end{array}
 \right\}.
\eq
We used the notation
\bq
 \hat{s}_{\alpha_1,...,\alpha_k}
 & = &
 \sum\limits_{i<j}
 \left( 2 p_{\alpha_i} p_{\alpha_j} + 2 \Delta_{\alpha_i \alpha_j} \right).
\eq
Let us mention that the coefficients $F_{w w'}$ are the ones appearing in the general BCJ relations for tree-level primitive
QCD amplitudes \cite{Bern:2008qj,Johansson:2015oia}.
We presented them here in a form which holds also for the massive case.
The general form of the BCJ relations is as follows:
Let $w_1=l_1 l_2 ... l_j$ be a sub-word, where particles of any type may occur 
and $w_2=l_{j+1} l_{j+2} ... l_{n-3}$ a second sub-word consisting only of gluon legs.
We further assume that $w = 1 w_1 (n-1) w_2 n \in W_2$.
The general BCJ relation reads
\bq
\label{general_BCJ}
 A_n\left( w \right)
 & = &
 \sum\limits_{w'} F_{w w'}
 A_n\left( w' \right).
\eq
As before, the sum is over all words occurring in
\bq
 1 \left( w_1 \shuffle S(w_2) \right) (n-1) n,
\eq
and the coefficients $F_{w w'}$ are defined as above.
The general BCJ relations of eq.~(\ref{general_BCJ}) follow from the fundamental BCJ relations \cite{Feng:2010my}.
}

\section{Comments on the rank of $F_{ww'}$}
\label{sect:upper_triangle_block}

We recall that the matrix $F_{ww'}$ is a $N_{\mathrm{basis}} \times N_{\mathrm{solutions}}$-dimensional matrix
with $N_{\mathrm{basis}} \le N_{\mathrm{solutions}}$, $w \in B$ and $w' \in B_{n_q \le 2}$.
The conjecture in eq.~(\ref{rank_condition_F}) states that the matrix $F$ has full row rank:
\bq
\label{appendix_rank_condition_F}
 \mathrm{rank} \; F_{w w'} & = & N_{\mathrm{basis}}.
\eq
In this appendix we show that in order to prove eq.~(\ref{appendix_rank_condition_F}) it is sufficient
to prove a weaker statement.
We first show that the matrix $F$ has an upper triangle block structure.
We do this by defining a suitable partial order for the elements of the basis $B$ and for the elements of the basis $B_{n_q \le 2}$.
A sufficient condition for eq.~(\ref{appendix_rank_condition_F}) is therefore that all (square) matrices on the main diagonal have full rank.

Let us start with $w \in B$. Let us write
\bq 
 w & = & 1 w_1 2 w_2 w_g (n-1) w_3 n,
\eq
with the condition that $w_2$ is either empty or ends with an antiquark and $w_g$ is either empty or contains only gluons.
This defines uniquely the sub-words $w_1$, $w_2$, $w_3$ and $w_g$.
The sub-words may be empty.
The sub-word $w_1$ encodes all particles which come after particle $1$ and before particle $2$ in the cyclic order,
the sub-word $w_3$ encodes all particles which come after particle $(n-1)$ and before particle $n$ in the cyclic order.
The sub-word $w_g$ encodes all gluons which directly precede particle $(n-1)$, 
the sub-word  $w_2$ encodes the remaining particles which come after particle $2$ and before particle $(n-1)$ in the cyclic
order.

Let us now look at the antiquarks in $w_3$. The corresponding quarks may either be in $w_3$ or in $w_1$.
They cannot be in $w_2$ (nor in $w_g$) since in this case they would have to cross the fermion line $2$-$(n-1)$.
We denote by $n_1$ the number of antiquarks in $w_3$, where the corresponding quark is again in $w_3$.
We denote by $n_2$ the number of antiquarks in $w_3$, where the corresponding quark is in $w_1$.
Furthermore we denote by $n_3$ the sum of the numbers of gluons in $w_3$ and $w_g$.
We associate to $w \in B$ the triple
\bq
\label{def_N_w}
 N(w) & = & \left( n_1, n_2, n_3 \right).
\eq
We define an order for these triples through
\bq
 \left( n_1', n_2', n_3' \right) & > & \left( n_1, n_2, n_3 \right)
\eq
if there is an $i$ such that $n_i'>n_i$ and $n_j'=n_j$ for all $j<i$.
This is just the lexicographical order for the triples $(n_1,n_2,n_3)$.
The triples $N(w)$ induce a partial order on $B$.

Let us now turn to $w' \in B_{n_q \le 2}$.
Let us write
\bq
 w' & = & 1 w_1' 2 w_2' (n-1) n.
\eq
Let us assume that
\bq
 w_2' & = & l_1' l_2' ... l_k'.
\eq
We now consider all possible splitting of $w_2'$ into two sub-words (with the empty words included)
\bq
 w_2' \;\; = \;\; u' v',
 & &
 u' \;\; = \;\; l_1' l_2' ... l_j',
 \;\;\;\;
 v' \;\; = \;\; l_{j+1}' ... l_k',
\eq
such that
\bq
\label{condition_basis_B_le_2}
 w & = & 1 w_1' 2 u' (n-1) v'^T n
\eq
is an element of $B$. There is either one or no possibility for such a splitting.
In the first case we set
\bq
 N'\left(w'\right) & = & N\left(w\right),
\eq
with $N(w)$ defined by eq.~(\ref{def_N_w}),
in the latter case we set
\bq
 N'\left(w'\right) & = & \left(-1,-1,-1\right)
\eq
This defines a partial order for $B_{n_q \le 2}$.
It is easy to see that there cannot be more than one possible splitting. Suppose $w_2'=u' v'$ is a possible 
splitting. Then $v'$ is either empty or must start with an antiquark.
All antiquarks in $u'$ have the standard orientation and do not cross other fermion lines, while all antiquarks
in $v'$ either have the opposite orientation or cross the fermion line $2$-$(n-1)$. These requirements make
the splitting unique.
If $N'(w')=(n_1,n_2,n_3) \neq (-1,-1,-1)$, then $n_1$ counts the number of antiquarks in $v'$ with
the opposite orientation, while $n_2$ counts the number of antiquarks in $v'$, 
which cross the fermion line $2$-$(n-1)$.
The variable $n_3$ gives the sum of the trailing gluons of $u'$ and the number of gluons in $v'$.

We may now order the basis $B$ by putting the elements $w$ with the highest $N(w)$ first.
In a similar way we order the basis $B_{n_q \le 2}$ by putting the elements $w'$ with the
highest $N'(w')$ first.
With respect to this ordering the matrix $F_{w w'}$ has an upper triangle block structure.
This means that 
\bq
\label{upper_triangle_block_structure}
 F_{w w'} = 0
 & & \mbox{if $N(w) < N'(w')$}.
\eq
Eq.~(\ref{upper_triangle_block_structure}) is easily understood as follows:
Let us consider a word $w=1 w_1 2 w_2 w_g (n-1) w_3 n \in B$ with $N(w)=(n_1,n_2,n_3)$.
The non-zero elements of the line $F_{w w'}$ with $w' \in B_{n_q \le 2}$ are the ones, where the letters
of the sub-word $w_3$ are inserted in arbitrary positions between the letters $1$ and $(n-1)$.
Suppose now that $w'=1 w_1' 2 u' v' (n-1) n$ with $N'(w')=(n_1',n_2',n_3')$
such that eq.~(\ref{condition_basis_B_le_2}) is satisfied.
The maximal number of antiquarks with the opposite orientation which may appear in $v'$ is exactly the
number $n_1$ of antiquarks in $w_3$, where the corresponding quark belongs also to $w_3$.
Thus we have $n_1' \le n_1$.

Let us now assume that $n_1'=n_1$. Then the maximal number of antiquarks appearing in $v'$ and
crossing the line $2$-$(n-1)$ is exactly (under the assumption $n_1'=n_1$) the number $n_2$ of antiquarks in $w_3$, where the corresponding
quark belongs to $w_1$.
Thus we have $n_2' \le n_2$.

Let us now look at the gluons. The maximal number of gluons appearing in $v'$ is exactly the number $n_3$
of gluons appearing in $w_3$ and $w_g$. 
Thus we have $n_3' \le n_3$.
This completes the proof of eq.~(\ref{upper_triangle_block_structure}).

Having established the upper triangle block structure 
it follows that a sufficient condition
for the matrix $F_{w w'}$  having full rank is the situation, 
where all the square matrices on the main diagonal have full rank.
In other words, we may consider the square sub-matrices
\bq
\label{condition_square_sub_matrices}
 F_{w w'}^{\mathrm{red}}
 & & 
 \mbox{with $N(w)=N'(w') \neq (-1,-1,-1)$}.
\eq
If for all sectors $N(w)=N'(w') \neq (-1,-1,-1)$ the corresponding sub-matrices $F_{w w'}^{\mathrm{red}}$ have
full rank, then it follows that $F_{w w'}$ has full rank.
Eq.~(\ref{condition_square_sub_matrices}) allows us to work with matrices of smaller dimensions and reduces therefore the complexity
of the problem.

We remark that for some sectors the matrix $F_{w w'}^{\mathrm{red}}$ has a diagonal block form and can be reduced further
to smaller square sub-matrices.
This is the case for sectors with $0 < n_1 < n_q-2$, where we may decompose $F_{w w'}^{\mathrm{red}}$ with respect
to the inequivalent antiquark flavour sets contributing to $n_1$.
Sectors with $0 < n_2 \le n_q-2$ decompose with respect to the ordered sequences of antiquarks in $w_3$ contributing to $n_2$
(and the corresponding reversed sequences in $v'$).
The sector $(n_1,n_2,n_3)=(0,0,0)$ contains all words $w$, which are at the same time 
elements of $B$ and $B_{n_q \le 2}$.
The matrix $F_{w w'}^{\mathrm{red}}$ for this sector is always the unit matrix.
However, the highest sector $(n_1,n_2,n_3)=(n_q-2,0,n_g)$ does in general not decompose further.

We have checked for all cases with $n \le 10$ external particles and generic external momenta that 
the corresponding matrices $F_{w w'}^{\mathrm{red}}$ have full rank.

\end{appendix}

\bibliography{/home/stefanw/notes/biblio}
\bibliographystyle{/home/stefanw/latex-style/h-physrev5}

\end{document}